%% file: main.tex
\documentclass[10pt,twocolumn,twoside]{IEEEtran}
\usepackage{amsmath,amsopn,amssymb,stackrel}
\usepackage{graphicx,xspace,color,soul}
\usepackage{epsfig,subfigure}
\usepackage{longtable,multirow}
\usepackage{array,float}
\usepackage{cite,citesort}
\usepackage{url}
\usepackage{algorithm,algorithmicx,algpseudocode}

\newcommand\ceil[1]{\lceil#1\rceil}

\usepackage{epstopdf}
\usepackage{stfloats}

\graphicspath{{figs/}}

\begin{document}
%\title{Markov Context Tree based Contour Coding}
\title{Context Tree based Image Contour Coding using A Geometric Prior}

\author{
Amin Zheng~\IEEEmembership{Student Member,~IEEE},
Gene Cheung~\IEEEmembership{Senior Member,~IEEE},
Dinei Florencio~\IEEEmembership{Fellow,~IEEE}
\begin{small}
\thanks{A. Zheng is with 
        Department of Electronic and Computer Engineering, 
        The Hong Kong University of Science and Technology, 
        Clear Water Bay, Hong Kong, China
        (e-mail: amzheng@connect.ust.hk).}
        \thanks{G. Cheung is with 
        National Institute of Informatics, 2-1-2, Hitotsubashi, Chiyoda-ku,
        Tokyo, Japan 101--8430 
        (e-mail: cheung@nii.ac.jp).}
        \thanks{D. Florencio is with  
        Microsoft Research,
        Redmond, WA USA 
        (e-mail: dinei@microsoft.com).}
%Phone Number: +1-778-782-7159 Fax Number: +1-778-782-4951
\end{small}
}%
\maketitle
\vspace{0.1in}

\begin{abstract}
If object contours in images are coded efficiently as side information, then they can facilitate advanced image / video coding techniques, such as graph Fourier transform coding or motion prediction of arbitrarily shaped pixel blocks. 
In this paper, we study the problem of lossless and lossy compression of detected contours in images. 
Specifically, we first convert a detected object contour composed of contiguous between-pixel edges to a sequence of directional symbols drawn from a small alphabet.
To encode the symbol sequence using arithmetic coding, we compute an optimal variable-length context tree (VCT) $\mathcal{T}$ via a maximum a posterior (MAP) formulation to estimate symbols' conditional probabilities. MAP prevents us from overfitting given a small training set $\mathcal{X}$ of past symbol sequences by identifying a VCT $\mathcal{T}$ that achieves a high likelihood $P(\mathcal{X}|\mathcal{T})$ of observing $\mathcal{X}$ given $\mathcal{T}$, and a large geometric prior $P(\mathcal{T})$ stating that image contours are more often straight than curvy.
For the lossy case, we design efficient dynamic programming (DP) algorithms that optimally trade off coding rate of an approximate contour $\hat{\mathbf{x}}$ given a VCT $\mathcal{T}$ with two notions of distortion of $\hat{\mathbf{x}}$ with respect to the original contour $\mathbf{x}$. 
To reduce the size of the DP tables, a total suffix tree is derived from a given VCT $\mathcal{T}$ for compact table entry indexing, reducing complexity. 
Experimental results show that for lossless contour coding, our proposed algorithm outperforms state-of-the-art context-based schemes consistently for both small and large training datasets. For lossy contour coding, our algorithms outperform comparable schemes in the literature in rate-distortion performance.
\end{abstract}


\IEEEpeerreviewmaketitle

%\vspace{-0.05in}
\section{Introduction}
\label{sec:intro}
\input{intro}

%\vspace{-0.1in}
\section{Related Work}
\label{sec:related}
\input{related}

%\vspace{-0.05in}
\section{Lossless Contour Coding}
\label{sec:lossless}
\input{lossless}

%\vspace{-0.05in}
\section{Lossy Contour Coding}
\label{sec:lossy}
\input{lossy}

\section{Starting Point Coding}
\label{sec:start}
\input{start}
\section{Experimental Results}
\label{sec:results}
\input{results}

%\vspace{-0.05in}
\section{Conclusion}
\label{sec:conclude}
\input{conclude}

\bibliographystyle{IEEEbib}
\bibliography{ref}

\end{document}

%% file: intro.tex
Advances in depth sensing technologies like Microsoft Kinect 2.0 mean that depth images---per pixel distances between physical objects in a 3D scene and the camera---can now be captured easily and inexpensively.
Depth imaging has in turn eased the detection of object contours in a captured image, which was traditionally a challenging computer vision problem \cite{grigorescu2003contour}.
If detected object contours in images are compressed efficiently as \textit{side information} (SI), then they can enable advanced image / video coding techniques such as \textit{graph Fourier transform} (GFT) coding \cite{shen10pcs,hu12icip,hu15,hu15spl} and motion prediction \cite{daribo12,daribo14} of arbitrarily shaped pixel blocks. 
Moreover, coded contours can be transmitted to a central cloud for computation-intensive object detection or activity recognition \cite{weinland2011survey}, at a much lower coding cost than compressed depth video.
We focus on the problem of coding object contours in this paper.

Object contour coding was studied extensively during 1990's when the concept of \textit{object-based video coding} (OBVC) was popular during the MPEG-4 video coding standardization.
However, the shape coding techniques developed to represent boundaries of objects \cite{katsaggelos1998mpeg}---called \textit{video object planes} (VOP)---were not efficient enough, resulting in large SI overhead that rendered OBVC uncompetitive in compression performance. 
Some contemporary interest in the problem has shifted to lossy contour coding, where curve- and spline-based approximations are used \cite{sohel2012sliding,lai2010arbitrary,zhu2014adaptive}.

Two recent technological trends have provided new ammunitions to revisit the old lossless contour coding problem.
First, advances in statistical machine learning have led to sophisticated algorithms that construct suitable \textit{variable-length context trees} (VCT) to estimate conditional probabilities given a training dataset of symbol strings \cite{begleiter2004prediction,rissanen1983universal,cleary1984data,ron1996power}. Estimated probability distributions can then be used to compress symbol strings of finite alphabets via arithmetic coding \cite{DCC1st1991}. 
Second, the availability of fast computing resource, locally or in a nearby cloudlet \cite{armbrust2010view}, allows this statistical learning on relevant training datasets to be performed in real-time in restricted cases. 
While promising, the convergence of these developments has led to a new \textit{``small data" statistical learning problem}:  due to either statistical non-stationarity and/or tight real-time application requirements, the size of dataset used for statistical training may be limited, and data overfitting becomes a significant concern. 

In response, in this paper we propose a \textit{maximum a posteriori} (MAP) formulation to optimize a VCT $\mathcal{T}$ for lossless and lossy contour coding. 
In particular, given small training data $\mathcal{X}$, we select a VCT $\mathcal{T}$ that has \textit{both} high likelihood $P(\mathcal{X} | \mathcal{T})$ (agreement between constructed contexts and observations), \textit{and} high prior probability $P(\mathcal{T})$---a geometric prior stating that image contours are more often straight than curvy.
Like \textit{Bayesian information criterion} (BIC) \cite{katz1981some}, we design the prior weight parameter $\alpha$ to scale naturally with training data size, so that when the volume of relevant training data becomes larger and thus more reliable, $\alpha$ becomes smaller.

For lossy coding, we design efficient dynamic programming (DP) algorithms that optimally trade off coding rate of an approximate contour $\hat{\mathbf{x}}$ given a VCT $\mathcal{T}$ with two different notions of distortion of $\hat{\mathbf{x}}$ with respect to the original contour $\mathbf{x}$: \textit{sum squared distance distortion} (SSDD) and \textit{maximum absolute distance distortion} (MADD) \cite{katsaggelos1998mpeg}. 
To reduce the size of the DP tables, a new tool called \textit{total suffix tree} (TST) is derived from a VCT $\mathcal{T}$ for compact table entry indexing, reducing computation complexity. 
Experimental results show that for lossless contour coding, our proposed algorithm outperforms state-of-the-art context-based schemes \cite{begleiter2004prediction,daribo14} consistently for both small and large training datasets. 
For lossy contour coding, our algorithms outperform comparable schemes in the literature \cite{katsaggelos1998mpeg,zhu2014adaptive} in rate-distortion (RD) performance.
Towards the goal of reproducible research \cite{vandewalle2009reproducible}, we have made the source code for lossless contour coding publicly available\footnote{\url{http://research.nii.ac.jp/~cheung/software.html}}.

The outline of the paper is as follows. We first discuss related works in Section \ref{sec:related}.
We then discuss our proposal for lossless and lossy contour coding in Section \ref{sec:lossless} and \ref{sec:lossy}, respectively.
We discuss efficient coding of contour starting points in Section \ref{sec:start}. 
Finally, we present experimental results and conclusion in Section \ref{sec:results} and \ref{sec:conclude}, respectively.

%% file: related.tex
%We discuss related work in lossless and lossy contour coding in order.

\subsection{Lossless Contour Coding}

Most works in lossless contour coding \cite{daribo14,freeman1978application,liu2005efficient,DCC1st1991,chan1995highly,estes1995efficient,turner1996efficient,egger1996region,jordan1998shape} first convert an image contour into a \textit{chain code} \cite{freeman1961}: a sequence of symbols each representing one of four or eight possible \textit{absolute} directions on the pixel grid. 
Alternatively, a \textit{differential chain code} (DCC) \cite{freeman1978application} that specifies \textit{relative} directions instead can be used. 
DCC symbols are entropy-coded using either Huffman \cite{liu2005efficient} or arithmetic coding \cite{DCC1st1991} given symbol probabilities. 
The challenge is to estimate conditional probabilities for DCC symbols given a set of training data; this is the core problem we address in Section \ref{sec:lossless}. 

\cite{daribo12icip,daribo14} propose a linear geometric model to estimate conditional probabilities of the next DCC symbol. 
In summary, given a window of previous edges, a line-of-best-fit that minimizes the sum of distances to the edges' endpoints is first constructed. Then the probability of a candidate direction for the next symbol is assumed inversely proportional to the angle difference between the direction and the fitted line. 
We show in Section \ref{sec:results} that this scheme is inferior in estimating symbol probabilities compared to context models,  because there are only a few possible angle differences for a small number of previous edges, limiting the expressiveness of the model.

%\red{still not clear what is the shortcomings of this scheme.}
%\blue{The conditional probability based on the geometric model is not accurate compared to that based on context model. The conditional probability is computed based on the angles between next edge and the predicted line direction. This mapping between angels and probability is not accurate. Beside, there are only a few fixed angels given a window of previous edges.}

%In most cases, the geometric model is able to find the most probable direction of the next chain code, but it fails to get the accurate probabilities of each direction. It is difficult to find a precise mapping between the probabilities and the angles.

An alternative approach is \textit{context modeling}: given a window of $l$ previous symbols (context) $\mathbf{x}^{i-1}_{i-l}$, compute the conditional probability $P(x_{i} | \mathbf{x}^{i-1}_{i-l})$ of the next symbol $x_{i}$ by counting the number of occurrences of $\mathbf{x}^{i-1}_{i-l}$ followed by $x_{i}$ in the training data. 
In \cite{kaneko1985encoding,DCC1st1991,chan1995highly,estes1995efficient,turner1996efficient}, Markov models of fixed order up to eight are used for lossless coding. 
However, in applications where the training data is small, there may be not enough occurrences of $\mathbf{x}^{i-1}_{i-l}$ to reliably estimate the conditional probabilities. 

VCT \cite{rissanen1983universal,begleiter2004prediction} provides a more flexible approach for Markov context modeling by allowing the context to have variable length. 
There are many ways to construct the VCT: Lempel-Ziv-78 (LZ78) \cite{ziv1977universal}, prediction by partial matching (PPM) \cite{cleary1984data}, and probabilistic suffix trees (PST) \cite{ron1996power}. 
LZ78 constructs a dictionary from scratch using the input data directly as training data. 
The probability estimation quality varies depending on the order of first appearing input symbols. 
%The later input symbols have more data to better estimate the probability. 
%\red{what do u mean by location of input data?} 
%\blue{Maybe we can show another drawback that for LZ78, since only input data is used for training, the training data is quite limited.}
PPM considers all contexts restricted by a maximum length with non-zero occurrences in the training data when building VCT. 
PPM is efficient for lossless sequential data compression if sufficient training data is available \cite{egger1996region,jordan1998shape}, but may suffer from overfitting if training data is limited. 

PST first constructs an initial VCT similar to PPM, and then the initial VCT is pruned using five user-selected thresholds \cite{ron1996power}. 
%\blue{Here, I want to emphasize that there are too many thresholds. I think we prune by a MAP formulation which is different from the threshold.}
PST algorithm is widely used for protein modeling problems \cite{buhlmann1999variable,bejerano2001variations}. 
Through pruning, PST algorithm can avoid overfitting in some cases. 
However, choosing properly the five application-specific thresholds used for pruning is difficult in general. 
In contrast, we propose a geometric shape prior---requiring only one parameter---to avoid overfitting specifically for contour coding.

%Most VCT-based compression methods use some training data to learn the model. \cite{akimov2007lossless,schiopu2013lossless} choose to use the test data to learn the context tree and the learned context tree is coded as side information. Training from test data can get true probability distribution of the context without overfitting problem, but the side information may consume lots of coding bits. In our proposed method for video based lossless contour coding, we use the contours in the previous frames as training data, which are available at both encoder and decoder.

\subsection{Lossy Contour Coding}

There are two main approaches to lossy contour coding: chain-code-based and vertex-based. In the first approach, the contours are first converted to chain codes as done in the lossless case. 
Then the chain codes are approximated according to different criteria before entropy coding. 
In \cite{yeh2002scalable}, an approximated chain code must be composed of several predefined sub-chain codes, and the indices of the sub-chain codes are coded. 
In \cite{zahir2007new}, a line processing technique generates straighter contour segments than the input, which are then efficiently compressed. 
In \cite{daribo14}, a chain code is simplified by removing ``irregularities", which are predefined non-smooth edge patterns. 
All these approaches approximate the chain codes without considering explicitly   distortion due to approximation, and one cannot specify a desired compression ratio. 
In contrast, we approximate a chain code via RD optimization algorithms, so that different compression ratios can be achieved.

Vertex-based approaches select representative points called vertices along the contour for coding. 
\cite{chung2000new,kuo2007new} use a top-down / bottom-up framework to select and code vertices independently. 
\cite{schuster1996rd,schuster1998operationally,katsaggelos1998mpeg} propose an operational rate-distortion (ORD) optimal algorithm to jointly select and code vertices. 
Many improvements on ORD have since been proposed, including \cite{sohel2006variable,sohel2007new,sohel2012sliding,lai2010arbitrary,zhu2014adaptive}. 
Compared to the chain-code-based approach, the vertex-based approach requires fewer points for coding, but each point (vertex) will consume more bits. We follow the chain-code-based approach for lossy coding in Section \ref{sec:lossy}.

%% file: lossless.tex
\begin{figure}[t]

\begin{minipage}[b]{.48\linewidth}
  \centering
  \centerline{\includegraphics[width=4cm]{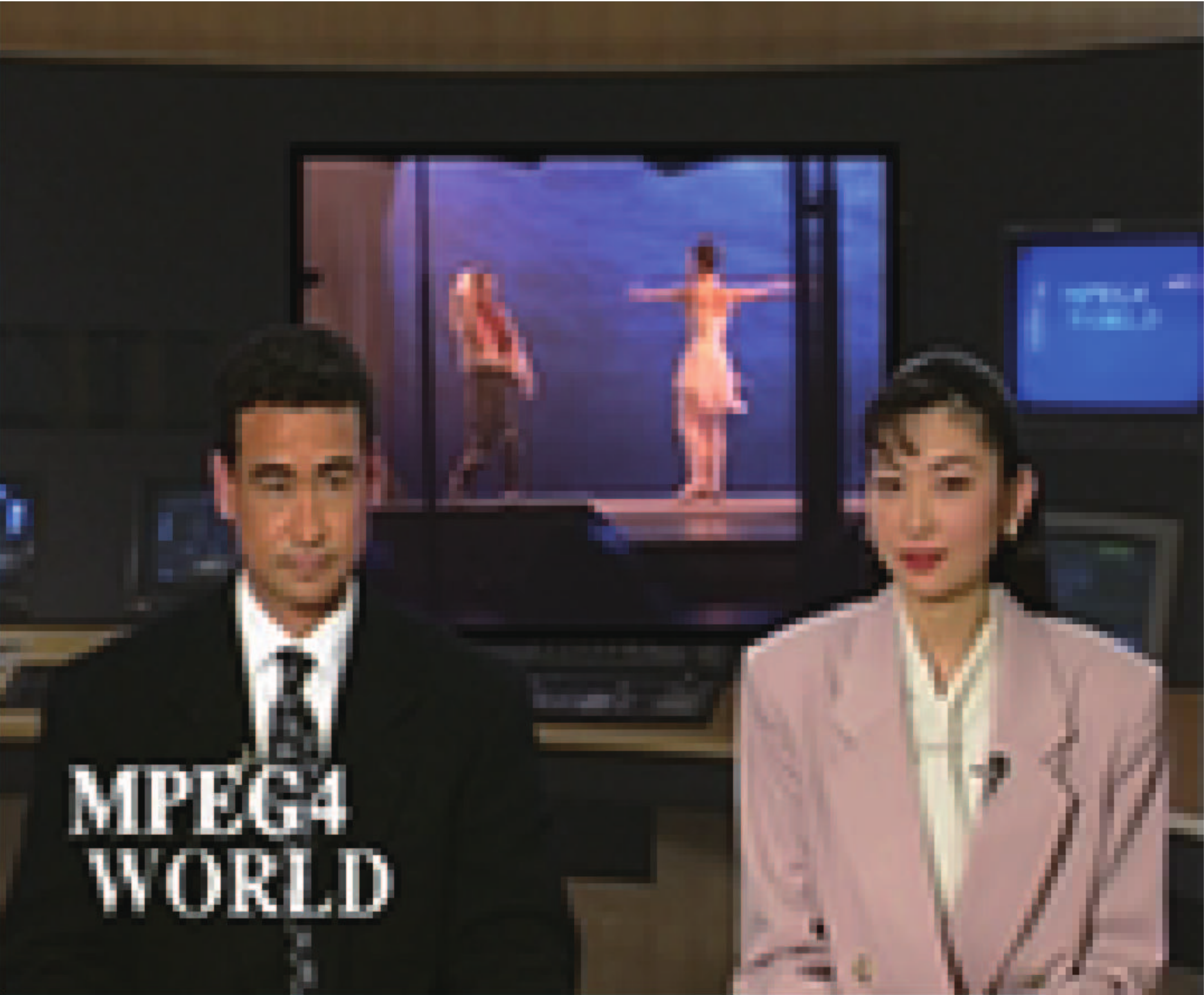}}
%  \vspace{1.5cm}
  \centerline{(a)}\medskip
\end{minipage}
\hfill
\begin{minipage}[b]{0.48\linewidth}
  \centering
  \centerline{\includegraphics[width=4cm]{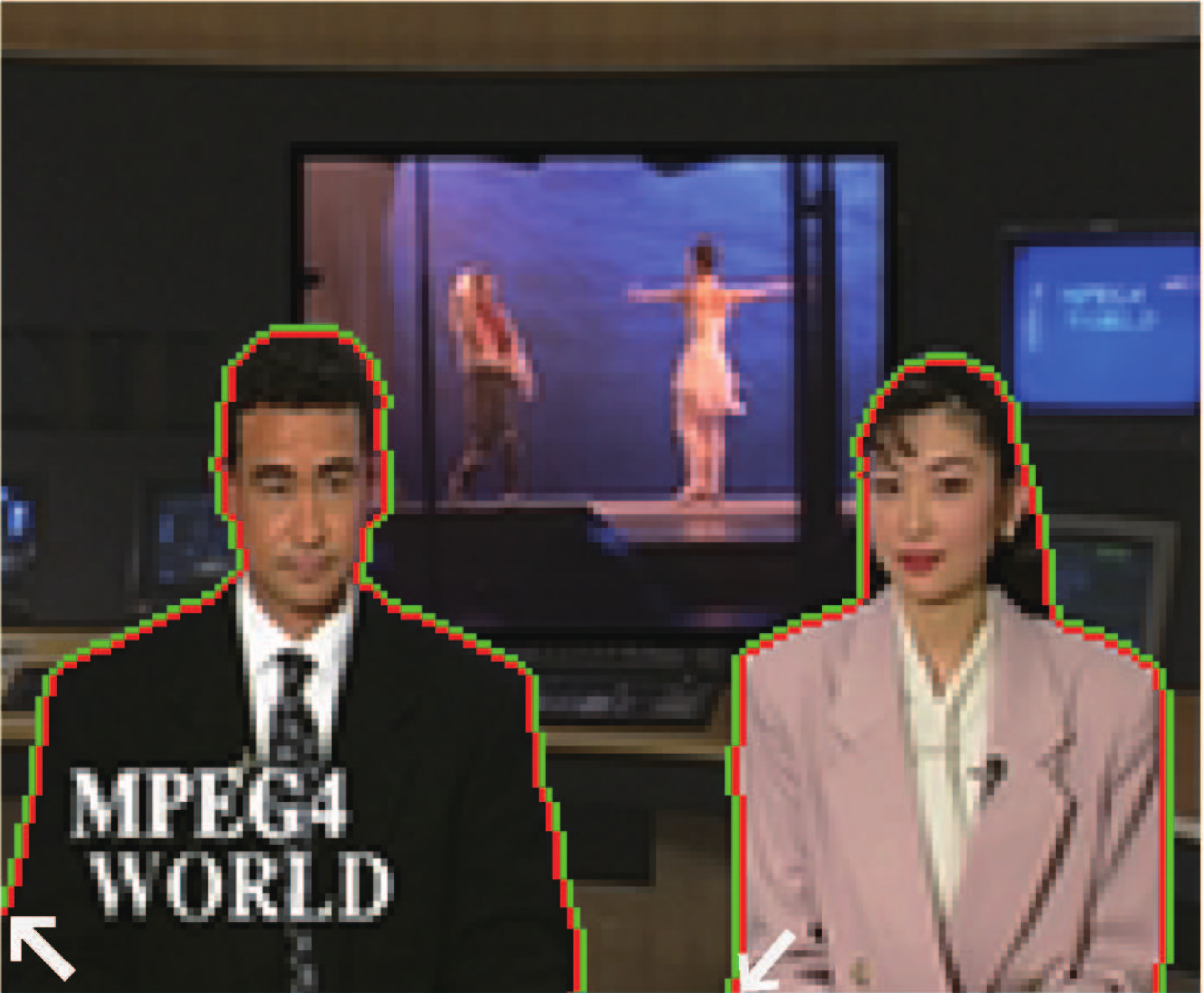}}
%  \vspace{1.5cm}
  \centerline{(b)}\medskip
\end{minipage}

\vspace{-0.1in}
\caption{An example of object contours of VOP in an MPEG4 video frame: (a) input image. (b) image with two object contours. The contours are the edges between the green and the red pixels. The starting points of the contours are indicated by the white arrows.}
\label{fig:vop}
\end{figure}

We first propose an algorithm to efficiently code object contours \textit{losslessly} in a target image. 
A small set of training images are used for statistical learning; we assume that the target image and the training images are statistically correlated, such as consecutive frames in a video sequence. 
We assume also that, as a pre-processing step, object contours in an image have first been either outlined manually, or detected automatically using an existing method such as gradient-based edge detection \cite{daribo14}. 
Each contour is defined by a starting point and a sequence of connected edges. 
See Fig.\;\ref{fig:vop} for two example contours of VOPs in a frame of MPEG4 test sequence \texttt{news}\footnote{\url{ftp://ftp.tnt.uni-hannover.de/pub/MPEG/mpeg4_masks/}}. 
Our objective here is to losslessly encode the sequence of connected edges in a given contour; coding of the starting points of contours is discussed later in Section \ref{sec:start}.

We first overview our lossless contour coding algorithm, based on the general concept of \textit{context tree} model \cite{rissanen1983universal,begleiter2004prediction}. The algorithm is composed of the following steps:
\begin{enumerate}
\item[1)] Convert each contour in a set of training images and a target image into a \textit{differential chain code} (DCC) \cite{freeman1978application}---a string of symbols each chosen from a size-three alphabet. 
\item[2)] Construct a suitable context tree given the DCC strings in the training images by solving a \textit{maximum a posterior} (MAP) problem.
\item[3)] For each symbol in an input DCC string (corresponding to a contour in the target image), identify the conditional probability distribution using the constructed context tree, which is used for arithmetic coding of the symbol.
\end{enumerate}
We discuss these steps in order.
Notations for the technical discussions in the sequel are shown in Table \ref{tab:notation}.

\begin{table}[tb]
\centering
\caption{Notations for lossless \& lossy contour coding}
\label{tab:notation}

\begin{center}

% Preview source code for paragraph 0

\begin{tabular}{c|c}
\hline 
\hline 
Notation & Description\tabularnewline
\hline 
$\mathcal{D}$ & alphabet of relative directions, $\mathcal{D} = \{\texttt{l}, \texttt{s}, \texttt{r}\}$\tabularnewline
$\mathcal{A}$ & alphabet of absolute directions, $\mathcal{A}=\{\texttt{N},\texttt{E},\texttt{S},\texttt{W}\}$\tabularnewline
%\hline 
$\mathcal{X}$ & training set of DCC strings \\
$\mathbf{x}(m)$ & $m$-th DCC string in training set $\mathcal{X}$ \\
$M$ & number of DCC strings in $\mathcal{X}$ \\
$l_m$ & number of symbols in string $\mathbf{x}(m)$ \\
$L$ & total number of symbols in $\mathcal{X}$ \\
$N(\mathbf{u})$ & number of occurrences of sub-string $\mathbf{u}$ in $\mathcal{X}$ \\
$P(x|\mathbf{u})$ & conditional probability of symbol $x$ given context $\mathbf{u}$ \\
%\hline 
$\mathbf{x},x_i$ & DCC string and its $i$-th symbol, $x_i\in \mathcal{D}$ \tabularnewline
$N$ & length of the DCC string\tabularnewline
$\hat{\mathbf{x}}$ & approximated DCC string \tabularnewline
$\hat{N}$ & length of the approximated DCC string \tabularnewline
% \hline 
% $x,v,s,z$ & One DCC symbol of $\texttt{l}$, $\texttt{s}$ and $\texttt{r}$\tabularnewline
% $\mathbf{u},\mathbf{v}$ & A sub-string of DCC symbols\tabularnewline
\multirow{2}{*}{$\mathbf{x}^{j}_{i}$} & A sub-string of length $j-i+1$ from $x_i$ to $x_j$, \tabularnewline
 & $\mathbf{x}^{j}_{i} = [{x}_{j},{x}_{j-1},\ldots,{x}_{i}]$\tabularnewline
%\hline 
$\mathbf{w}$ & context \tabularnewline
$\mathcal{T},\mathcal{T}^{\ast}$ & context tree, an optimal context tree \tabularnewline
$\mathcal{F}$ & context forest composed of all possible context trees \tabularnewline
$D$ & maximum depth of the context tree \tabularnewline
%\hline 
$\mathcal{T}^{0}$ & initial context tree \tabularnewline
$K$ & Number of nodes on $\mathcal{T}^{0}$ \tabularnewline
%\hline 
$\mathcal{T}_s$ & total suffix tree\tabularnewline
%\hline 
$\alpha$ & prior weight parameter, $\alpha = a\ln L$\tabularnewline
%\hline 
$D_{\max}$ & maximum distortion of the approximated DCC string\tabularnewline
\hline 
\hline 
\end{tabular}

\end{center}
\end{table}

\subsection{Differential Chain Code}
\label{subesc:differentialChainCode}

\begin{figure}[tb]

\begin{minipage}[b]{.48\linewidth}
  \centering
  \centerline{\includegraphics[width=4cm]{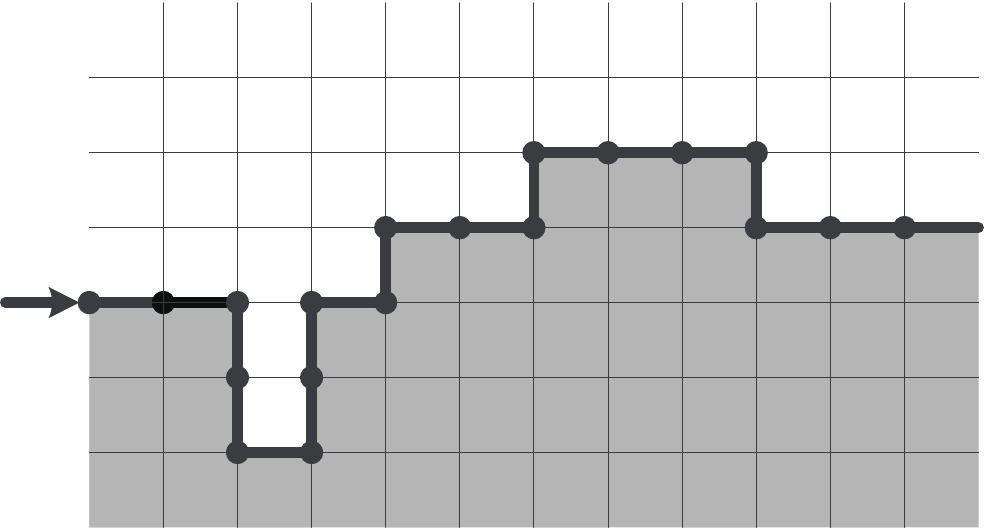}}
%  \vspace{1.5cm}
  \centerline{(a)}\medskip
\end{minipage}
\hfill
\begin{minipage}[b]{0.48\linewidth}
  \centering
  \centerline{\includegraphics[width=3cm]{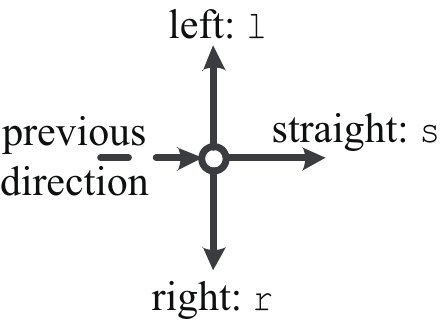}}
%  \vspace{1.5cm}
  \centerline{(b)}\medskip
\end{minipage}

\vspace{-0.3cm}
\caption{(a) An example of a contour represented by a four-connected chain codes: $\texttt{east}\!-\!\texttt{s}\!-\!\texttt{r}\!-\!\texttt{s}-\!\texttt{l}\!-\!\texttt{l}\!-\!\texttt{s}-\!\texttt{r}\!-\!\texttt{l}\!-\!\texttt{r}\!-\!\texttt{s}\!-\!\texttt{l}\!-\!\texttt{r}\!-\!\texttt{s}\!-\!\texttt{s}\!-\!\texttt{r}\!-\!\texttt{l}\!-\!\texttt{s}\!-\!\texttt{s}$. (b) directional code.}
\label{fig:contour_definition}
\end{figure}

We first convert each contour into a DCC string; DCC is one member of the family of chain codes proposed by Freeman \cite{freeman1978application}.
A contour is composed of a sequence of ``between-pixel" edges that divide pixels in the local neighborhood into two sides, as illustrated in Fig.\;\ref{fig:contour_definition}(a).  
A length-$(N+1)$ contour can be compactly described as a symbol string denoted by $\mathbf{x}^o = [x_0, \ldots, x_{N}]$. For the first edge $x_0$, we assume equal probability for each of four \textit{absolute} directions, \texttt{north}, \texttt{east}, \texttt{south} and \texttt{west}, with respect to the starting point. 
For each subsequent DCC symbol $x_i$, $i \geq 1$, only three \textit{relative} directions are possible on a 2D grid with respect to the previous symbol $x_{i-1}$: \texttt{left}, \texttt{straight} and \texttt{right}, as shown in Fig.\;\ref{fig:contour_definition}(b). 
We denote them by \texttt{l}, \texttt{s} and \texttt{r}, respectively, which constitute a size-three alphabet $\mathcal{D} = \{\texttt{l}, \texttt{s}, \texttt{r}\}$. 
The problem is thus to code a DCC string $\mathbf{x}$ (without the first edge), where $x_i \in \mathcal{D}$ for $i \geq 1$.

\subsection{Definition of Context Tree}
\label{subsec:definitionOfCT}
 
%\red{if space permits, it'd be nice to have a table of notations.}

We first define notations. 
Denote by $\mathbf{x}(m)$, $1 \leq m \leq M$, the $m$-th DCC string in the training set $\mathcal{X} = \{\mathbf{x}(1),\ldots,\mathbf{x}(M)\}$, where $M$ denotes the total number of DCC strings in $\mathcal{X}$. The length of $\mathbf{x}(m)$ is denoted by $l_m$, and the total number of symbols in $\mathcal{X}$ is denoted by $L = \sum^{M}_{m=1}l_m$. 

Denote by $\mathbf{x}^{j}_{i} = [{x}_{j},{x}_{j-1},\ldots,{x}_{i}]$, $i<j$ and $i,j \in \mathbb{Z}^+$, a \textit{sub-string} of length $j-i+1$ from the $i$-th symbol $x_i$ to the $j$-th symbol $x_j$ in reverse order. Further, denote by $\mathbf{u}\mathbf{v}$ the concatenation of sub-strings $\mathbf{u}$ and $\mathbf{v}$.

We now define $N(\mathbf{u})$ as the number of occurrences of sub-string $\mathbf{u}$ in the training set $\mathcal{X}$. $N(\mathbf{u})$ can be computed as:
\begin{equation}
N(\mathbf{u})=\sum^{M}_{m=1}\sum_{i=1}^{l_m - |\mathbf{u}| + 1}
%1_{[\mathbf{x}(j)^{i+|\mathbf{u}|-1}_{i}=\mathbf{u}]} 
\mathbf{1} \left(
\mathbf{x}(m)_i^{i+|\mathbf{u}|-1} = \mathbf{u}
\right) 
\end{equation}
%where $\mathbf{u} \in \stackrel [k=1] {+\infty}{\cup} \mathcal{D}^{k}$. 
where $\mathbf{1}(\mathbf{c})$ is an indicator function that evaluates to $1$ if the specified binary clause $\mathbf{c}$ is true and $0$ otherwise.

Denote by $P(x|\mathbf{u})$ the conditional probability of symbol $x$ occurring given its previous sub-string is $\mathbf{u}$, where $x\in\mathcal{D}$. Given training data $\mathcal{X}$, $P(x|\mathbf{u})$ can be estimated using $N(\mathbf{u})$ as done in \cite{buhlmann1999variable},
\begin{equation}
\begin{array}{cc}
P(x|\mathbf{u})=\frac{N(x\mathbf{u})}{N(\mathbf{u})}
\end{array}
\label{eq:cal_prob}
\end{equation}

%\subsubsection{Context Tree}
%\label{subsubsec:context_tree}

Given $\mathcal{X}$, we learn a context model to assign a conditional probability to any symbol given its previous symbols in a DCC string. Specifically, to calculate the conditional probability $P(x_i|\mathbf{x}^{i-1}_{1})$ for the symbol $x_i$ given all its previous symbols $\mathbf{x}^{i-1}_{1}$, the model determines a \textit{context} $\mathbf{w}$ to calculate $P(x_i|\mathbf{w})$, where $\mathbf{w}$ is a \textit{prefix} of the sub-string $\mathbf{x}^{i-1}_{1}$, \textit{i.e.}, $\mathbf{w} = \mathbf{x}^{i-1}_{i-l}$ for some context length $l$:
\begin{equation}
P(x_i|\mathbf{x}^{i-1}_{1})=P(x_i|\mathbf{w})
\label{eq:conditionalProbContext}
\end{equation}
$P(x_i|\mathbf{w})$ is calculated using (\ref{eq:cal_prob}) given $\mathcal{X}$. 
The context model determines a unique context $\mathbf{w}$ of finite length for every possible past $\mathbf{x}^{i-1}_{1}$. The set of all mappings from $\mathbf{x}^{i-1}_{1}$ to $\mathbf{w}$ can be represented compactly as a context tree.

\begin{figure}[t]

\begin{minipage}[b]{1\linewidth}
  \centering
  \centerline{\includegraphics[width=8 cm]{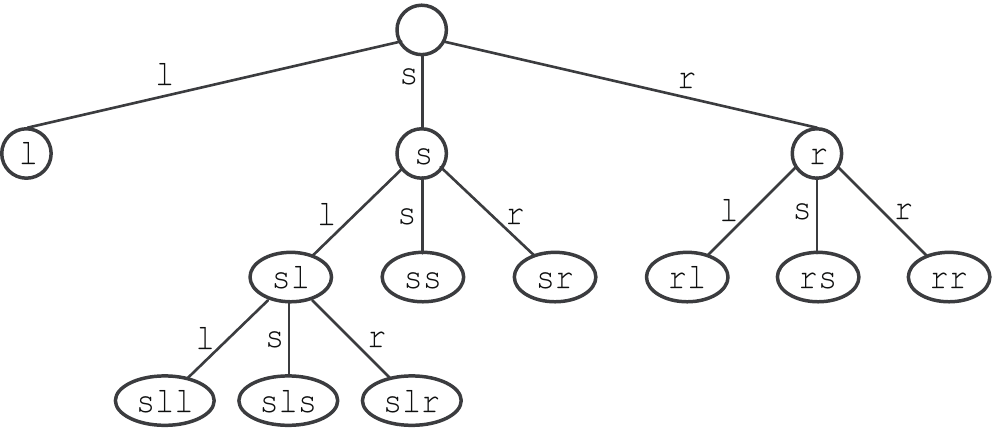}}
  %\vspace{0.1cm}
  \centerline{}
  %\medskip
\end{minipage}

\vspace{-0.2cm}
\caption{An example of context tree. Each node is a sub-string and the root node is an empty sub-string. The contexts are all the end nodes on $\mathcal{T}$: $\mathcal{T}=\{\texttt{l},\texttt{sll},\texttt{sls},\texttt{slr},\texttt{ss},\texttt{sr},\texttt{rl},\texttt{rs},\texttt{rr}\}$.}

\label{fig:MCT_statespace}
\end{figure}

Denote by $\mathcal{T}$ the context tree, where $\mathcal{T}$ is a \textit{full ternary tree}: each node has either zero children (an \textit{end node}) or all three children (an \textit{intermediate node}). 
The root node has an empty sub-string, and each child node has a sub-string $\mathbf{u}   x$ that is a concatenation of: i) its parent's sub-string $\mathbf{u}$ if any, and ii) the symbol $x$ (one of \texttt{l}, \texttt{s} and \texttt{r}) representing the link connecting the parent node and itself in $\mathcal{T}$. 
An example is shown in Fig.\;\ref{fig:MCT_statespace}. 
The sub-strings of the end nodes are the contexts of the tree $\mathcal{T}$. 
%Note that $\mathcal{T}$ is also denoted as the set of all the contexts on the tree. 
For each $\mathbf{x}^{i-1}_{1}$, a context $\mathbf{w}$ is obtained by traversing $\mathcal{T}$ from the root node until an end node, matching symbols $x_{i-1}, x_{i-2}, \ldots$ into the past.

All possible context trees constitute a \textit{context forest} denoted by $\mathcal{F}$. The modeling problem is to find the best tree $\mathcal{T}$ from the forest $\mathcal{F}$ given $\mathcal{X}$.

\subsection{Problem Definition for Optimal Context Tree}
\label{subsec:formulationOfMCT}

Given limited observations $\mathcal{X}$, we aim to find a suitable tree $\mathcal{T}$ that best describes $\mathcal{X}$ without overfitting. 
We first write the posterior probability of $\mathcal{T}$ given $\mathcal{X}$ via Bayes' rule:
\begin{equation}
P(\mathcal{T}|\mathcal{X})=\frac{P(\mathcal{X}|\mathcal{T})P(\mathcal{T})}{P(\mathcal{X})}
\label{eq:bayes_rules}
\end{equation}
where $P(\mathcal{X}|\mathcal{T})$ is the likelihood of observing $\mathcal{X}$ given context tree $\mathcal{T}$, and $P(\mathcal{T})$ is the prior which describes \textit{a priori} knowledge about the context tree. We next define the likelihood and prior terms, and then use the MAP estimator to formulate the context tree optimization problem. 

%\begin{figure*}[!b]
%
%% IEEE uses as a separator
%\hrulefill
%% The spacer can be tweaked to stop underfull vboxes.
%
%% ensure that we have normalsize text
%\normalsize
%% Store the current equation number.
%\setcounter{mytempeqncnt}{\value{equation}}
%% Set the equation number to one less than the one
%% desired for the first equation here.
%% The value here will have to changed if equations
%% are added or removed prior to the place these
%% equations are referenced in the main text.
%\setcounter{equation}{6}
%
%\begin{equation}
%\label{eq:sw}
%s(\mathbf{w})=\underset{1 \leq k \leq |\mathbf{w}|+1}{\max}\left\{\frac{|(p_{k,x}-p_{1,x})\cdot(p_{|\mathbf{w}|+1,y}-p_{1,y})-(p_{k,y}- p_{1,y})\cdot(p_{|\mathbf{w}|+1,x}-p_{1,x}) |}{\sqrt{(p_{|\mathbf{w}|+1,x}-p_{1,x})^2+(p_{|\mathbf{w}|+1,y}-p_{1,y})^2}}  \right\}
%\end{equation}
%
%% Restore the current equation number.
%\setcounter{equation}{\value{mytempeqncnt}}
%
%\vspace*{4pt}
%
%\end{figure*}

\subsubsection{Likelihood Term}
\label{subsubsec:likelihood}

The likelihood is defined as the joint conditional probability of all the observed symbols in $\mathcal{X}$ given their past and $\mathcal{T}$,
\begin{equation}
P(\mathcal{X}|\mathcal{T}) = \prod_{m=1}^{M}\;\prod_{i=1}^{l_m} P(x(m)_i|\mathbf{x}(m)^{i-1}_1, \mathcal{T})
\label{eq:joint_conditional_distribution}
\end{equation}
Given tree $\mathcal{T}$, for each symbol $x(m)_i$ a prefix (context) $\mathbf{w}$ of past symbols $\mathbf{x}(m)^{i-1}_1$ is identified to compute the conditional probability. 
Hence (\ref{eq:joint_conditional_distribution}) can be rewritten as follows using (\ref{eq:conditionalProbContext}), similarly done in \cite{alcaraz2010bi}:
\begin{equation}
P(\mathcal{X}|\mathcal{T}) = \underset{\mathbf{w}\in \mathcal{T}}{\prod}\;\underset{x\in\mathcal{D}}{\prod}P(x|\mathbf{w})^{N(x\mathbf{w})}
\label{eq:likelihood}
\end{equation}

\subsubsection{Prior Term}
\label{subsubsec:prior}

Overfitting occurs if the complexity of a model is too large for the given observed data size. 
In the case of context tree, it means that the number of occurrences $N(\mathbf{u})$ of a particular context $\mathbf{u}$ is too small to have probabilities $P(x|\mathbf{u})$ reliably estimated using $\frac{N(x\mathbf{u})}{N(\mathbf{u})}$. 
To avoid overfitting, one can design a prior to control the size of $\mathcal{T}$---the complexity of the model---depending on the volume of training data. 
This general idea is used for example in \cite{alcaraz2010bi}, where \textit{Bayesian information criterion} (BIC) \cite{katz1981some} is employed to constrain the order of a fixed-length Markov model for given data size.
%\red{check if the rewritten text is correct.}
%\blue{It is not precise. \cite{alcaraz2010bi} only uses fixed length Markov model. It selects the maximum order (fixed length) using BIC.}

%Different from the prior in BIC which treats each parameter equally, our proposed geometric prior penalizes the context by $s(\mathbf{w})$, which prefers straighter context than curvy context.

% The previous equation was number six.
% Account for the double column equations here.
% \addtocounter{equation}{1}

\begin{figure}[h]

\begin{minipage}[b]{.32\linewidth}
  \centering
  \centerline{\includegraphics[width=1.8cm]{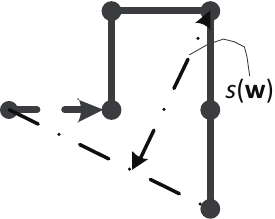}}
%  \vspace{1.5cm}
  \centerline{(a)}\medskip
\end{minipage}
\hfill
\begin{minipage}[b]{0.32\linewidth}
  \centering
  \centerline{\includegraphics[width=1.3cm]{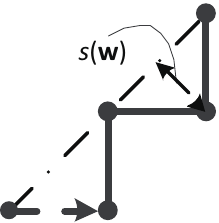}}
%  \vspace{1.5cm}
  \centerline{(b)}\medskip
\end{minipage}
\hfill
\begin{minipage}[b]{0.32\linewidth}
  \centering
  \centerline{\includegraphics[width=1.8cm]{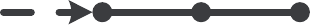}}
%  \vspace{1.5cm}
  \centerline{(c)}\medskip
\end{minipage}

%\vspace{-0.3cm}
\caption{Three examples of $s(\mathbf{w})$. (a) $\mathbf{w}=\texttt{srrl}$ and $s(\mathbf{w})=4\sqrt{5}/{5}$. (b) $\mathbf{w}=\texttt{lrl}$ and $s(\mathbf{w})={\sqrt{2}}/{2}$. (c) $\mathbf{w}=\texttt{ss}$ and $s(\mathbf{w})=0$.}
\label{fig:prior}
\end{figure}

In the case of contour coding, we propose a \textit{geometric prior}, defined as the sum of \textit{straightness}, $s(\mathbf{w})$, of all contexts $\mathbf{w}$ in tree $\mathcal{T}$, based on the assumption that contours in natural images are more likely straight than curvy. 
We calculate $s(\mathbf{w})$ as follows. 
We first map a context $\mathbf{w}$ to a \textit{shape segment} on a 2D grid with $|\mathbf{w}|+2$ points from the most recent symbol $w_{|\mathbf{w}|}$ to the symbol $w_1$ furthest in the past \textit{plus} an initial edge. 
Without loss of generality, we assume that the absolute direction of the initial edge is \texttt{East}. 
As an example, Fig.\;\ref{fig:prior}(b) shows a context $\mathbf{w}=\texttt{lrl}$ with five points, where the dotted arrow is the initial edge. 
%After turning left, right and left in orders, we get a shape segment with five points.   
Denote by $\mathbf{p}_k$, $1\leq k \leq |\mathbf{w}|+2$, the 2D coordinate of the $k$-th point. 
$s(\mathbf{w})$ is defined as the \textit{maximum distance} $\mathrm{dist}(\,)$ from any $\mathbf{p}_k$ to a straight line $f(\mathbf{p}_1, \mathbf{p}_{|\mathbf{w}|+2})$ connecting $\mathbf{p}_1$ and $\mathbf{p}_{|\mathbf{w}|+2}$, 
\begin{equation}
s(\mathbf{w}) = \max_k 
\mathrm{dist} \left(\mathbf{p}_k, \, f(\mathbf{p}_1, \mathbf{p}_{|\mathbf{w}|+2}) \right)
\end{equation}
Some examples of $s(\mathbf{w})$ are shown in Fig.\;\ref{fig:prior}.

%as mathematically defined in (\ref{eq:sw}). 
%Note that $w_{|\mathbf{w}|}$ only defines a relative direction to its previous symbol. 
%The absolute direction of $w_{\mathbf{w}}$ does not affect the calculation of $s(\mathbf{w})$. Without loss of generality, we set the coordinates of $\mathbf{p}_1$ and $\mathbf{p}_2$ defined by $w_{|\mathbf{w}|}$ to $[0,0]$ and $[1,0]$, respectively. 

We can now define prior $P(\mathcal{T})$ based on the sum of $s(\mathbf{w})$ of all contexts $\mathbf{w}$ in $\mathcal{T}$ as follows:
\begin{equation}
P(\mathcal{T})=\exp\left(-{\alpha}\sum\limits_{\mathbf{w}\in\mathcal{T}}s({\mathbf{w}})\right)
\label{eq:prior}
\end{equation}
%Here, we use Laplacian function not Gaussian function for efficient computation. 
where $\alpha$ is an important parameter to be discussed soon.  
We see that in general a tree $\mathcal{T}$ with fewer contexts $\mathbf{w}$ has a smaller sum in (\ref{eq:prior}), meaning that our prior tends to reduce the number of parameters. 
Further, in general fewer contexts also means a shallower context tree. Thus, our prior promotes shorter contexts, which is also reasonable.

%Finally, since the contour detection process may induce noises which usually curve the contours, our proposed geometric prior can avoid these noises by preferring to the straighter context.

\subsubsection{MAP Estimation}
\label{subsubsec:MAP}

We can now write the optimal context tree problem via a MAP formulation:
\begin{equation}
\mathcal{T}^{\ast} = \underset{\mathcal{T}\in \mathcal{F}}{\arg \max}\ P(\mathcal{X}|\mathcal{T})P(\mathcal{T})
\label{eq:MAP_initial}
\end{equation}
%The denominator $P(\mathcal{X})$ in (\ref{eq:bayes_rules}) does not depend on $\mathcal{T}$ and therefor plays no role in the optimization. 

(\ref{eq:MAP_initial}) can be rewritten as: 
\begin{equation}
\begin{split}
&\mathcal{T}^{\ast}=\underset{\mathcal{T} \in \mathcal{F}}{\arg \max} \\
&\left\{\underset{\mathbf{w}\in\mathcal{T}}{\prod} \underset{x\in\mathcal{D}}{\prod}P(x|\mathbf{w})^{N(x\mathbf{w})} \cdot  \exp\left(-{\alpha}\sum\limits_{\mathbf{w}\in\mathcal{T}}s({\mathbf{w}})\right) \right\}
\end{split}
\label{eq:Best_MCT_Definition}
\end{equation}

For ease of computation, we minimize the negative log of (\ref{eq:Best_MCT_Definition}) instead and divide by $L$:
\begin{equation}
\begin{split}
&F(\mathcal{T})=\\
&-\frac{1}{L}\underset{\mathbf{w}\in\mathcal{T}}{\sum}\underset{x\in\mathcal{D}}{\sum}N(x\mathbf{w})\cdot\ln{P(x|\mathbf{w})}+\frac{\alpha}{L}\sum\limits_{\mathbf{w}\in\mathcal{T}}s({\mathbf{w}})
\end{split}
\label{eq:jmap_ln}
\end{equation} 
The first term in (\ref{eq:jmap_ln}) can be interpreted as the average information of all symbols in observed $\mathcal{X}$. The second term is the average straightness of all contexts. $\alpha$ weighs the relative importance of the prior against the likelihood.

\subsubsection{Selection of Weighting Parameter}

Similar in principle to BIC \cite{katz1981some}, we define $\alpha = a \ln{L}$, where $a$ is a chosen parameter. 
In doing so, when the size of the training data $L$ becomes larger, the weight factor for the prior $a\ln{L}/L$ becomes smaller.
This agrees with the intuition that when sufficient training data is available, the prior term becomes less important.

%Particularly, when $L$ is large enough, the objective will be mostly decided by the likelihood. In summary, the weight between the likelihood and the prior is adapted to $L$, which works for both the small data and large data applications. 

\subsection{Optimization of Context Tree}
\label{subsec:optimizationOfCT}

The optimization of context tree consists of two main steps. We first construct an initial context tree denoted by $\mathcal{T}^{0}$ by collecting statistics from $\mathcal{X}$. We then prune $\mathcal{T}^{0}$ to get the optimal context tree $\mathcal{T}^{\ast}$ by minimizing the objective (\ref{eq:jmap_ln}).

\subsubsection{Construction of Initial Context Tree}
\label{subsubsec:constructionOfInitialContextTree}

Since $\mathcal{T}^{\ast}$ is a sub-tree of $\mathcal{T}^{0}$, each node (intermediate node or end node) in $\mathcal{T}^{0}$ can be a potential context (end node) in $\mathcal{T}^{\ast}$.
Given a maximum tree depth $D$, \cite{buhlmann1999variable,akimov2007lossless} construct $\mathcal{T}^{0}$ by collecting statistics for all nodes, \textit{i.e.}, $(3^{D+1}-1)/2$, which means that the required memory is exponential with respect to $D$.
To reduce the memory requirement, we enforce an upper-bound $K$ on the number of nodes in $\mathcal{T}^{0}$ given $D$.
Specifically, we first choose $D=\ceil{\ln{L}/\ln{3}}$ as done in \cite{rissanen1983universal}, which ensures a large enough $D$ to capture natural statistics of the training data of length $L$. 
We then choose $K=3 D^3$ in our experiments, which is much smaller than $(3^{D+1}-1)/2$.
%\red{the choice of $D$ should be moved earlier when it first appears. further, you still don't say how $K$ is selected. I think u meant to say $D$ is selected large enough to capture the long but statistically reliable contexts, while $K$ is selected to limit memory size. right??}
%\red{should the discussion on how $K$ is selected given $D$ be moved here?}
%\blue{Did you mean ``choosing K potential contexts from 2K potential contexts''?}
%\red{no, I mean how the value $K$ is chosen. 15? 500? we need to show $K$ is not a number that is proportional to $3^D$.}
%\blue{I have added the selection of $K$.}

Having chosen $D$ and $K$, we parse the training data $\mathcal{X}$ to collect statistics for $K$ potential contexts.
% To avoid choosing context $\mathbf{u}$ with inadequate number of occurrences $N(\mathbf{u})$, 
Specifically, we traverse $\mathcal{X}$ once to first collect statistics for the first $2K$ different sub-strings $\mathbf{u}$ we encounter, where $|\mathbf{u}|\leq D$. 
Each sub-string $\mathbf{u}$ has three counters which store the number of occurrences of sub-strings $l\mathbf{u}$, $s\mathbf{u}$ and $r\mathbf{u}$, \textit{i.e.}, $N(l\mathbf{u})$, $N(s\mathbf{u})$ and $N(r\mathbf{u})$.
Then we keep only the $K$ sub-strings with the largest numbers of occurrences to construct $\mathcal{T}^{0}$, as described in Algorithm \ref{al:initialContextTree}.

\begin{algorithm}
\caption{Choose $K$ potential contexts}
\label{al:initialContextTree}
\begin{algorithmic}[1]

\State{Initialize $\mathcal{T}^{0}$ to an empty tree with only root node}

\For{each symbol $x_i,\;i\geq D+2$ in $\mathcal{X}$, match $x_{i-k}^{i-1}$ with nodes on $\mathcal{T}^{0}$ from $k=1$ to $k=D$ in order}

\If{there exist a node $\mathbf{u}=x_{i-k}^{i-1}$ on $\mathcal{T}^{0}$}
	\State{increase the counter $N(x_i\mathbf{u})$ by $1$}
\ElsIf{number of nodes on $\mathcal{T}^{0}$ is less than $2K$}
	\State{add node $\mathbf{u}=x_{i-k}^{i-1}$ to $\mathcal{T}^{0}$}
\EndIf

\EndFor

\State{Sort nodes on $\mathcal{T}^{0}$ by $N(\mathbf{u})$ in descending order and choose the first $K$ nodes from the sorted nodes.}

\end{algorithmic}
\end{algorithm}

The obtained tree by Algorithm\;\ref{al:initialContextTree} may not be a full tree, because some intermediate nodes may have only one or two children.
Thus, we add children to these intermediate nodes to ensure that each intermediate node has three children.
We assume that the statistics of a newly added child $\mathbf{u}v$ is the same as its parent $\mathbf{u}$; 
%this ensures that this added child does not alter the computed likelihood term in (\ref{eq:likelihood}).
\textit{i.e.}, $P(x|\mathbf{u}v)=P(x|\mathbf{u})$.
The occurrences of the added children are set to ensure that the total number of occurrences of all three children is equal to the number of occurrence of their parent.
Specifically, $N(\mathbf{u}v)=N(\mathbf{u})-\underset{z\in \mathcal{D},z\neq v}{\sum}N(\mathbf{u}z)$ if only one child $\mathbf{u} v$ is added, and $N(\mathbf{u}v)=N(\mathbf{u}s)=\frac{1}{2}(N(\mathbf{u})-\underset{z\in \mathcal{D},z\neq v, z\neq s}{\sum}N(\mathbf{u}z))$ if two children $\mathbf{u} v$ and $\mathbf{u} s$ are added.
 
%\red{this is still not complete. what if the u r adding two children to an intermediate node initially with only one child?}
%\blue{I have put it in another way.}

After the adding procedure, we arrive at an initial context tree $\mathcal{T}^0$ with maximum depth $D$.
The memory requirement of the initialization is $O(3\cdot 2K) = O(K)$, and the time complexity is $O(K\cdot L)$.

%\begin{figure}[h]
%
%\begin{minipage}[b]{1\linewidth}
%  \centering
%  \centerline{\includegraphics[width=4 cm]{sub_MCT}}
%  %\vspace{0.1cm}
%  \centerline{}
%  %\medskip
%\end{minipage}
%
%\vspace{-0.2cm}
%\caption{An example of context sub-tree rooted at node $sl$ of the context tree shown in Fig.\;\ref{fig:MCT_statespace}. The contexts are the leaves on the context sub-tree: $\mathcal{T}_{sl}=\{sll,sls,slr\}$. \red{what is new in this figure that is not already shown in realier Fig. 3?}}
%\label{fig:sub_MCT}
%\end{figure}

\subsubsection{Pruning of Initial Context Tree}
\label{subsubsec:optimalCT}

The obtained initial tree $\mathcal{T}^{0}$ is then pruned to minimize the objective (\ref{eq:jmap_ln}), resulting in an optimal tree $\mathcal{T}^{\ast}$.
Since both the likelihood and the prior in (\ref{eq:jmap_ln}) are summations of all the contexts (end nodes) in $\mathcal{T}$, we rewrite the objective as the sum of \textit{end node cost} denoted by $f(\mathbf{w})$:
\begin{equation}
F(\mathcal{T})=\underset{\mathbf{w}\in\mathcal{T}}{\sum} f(\mathbf{w})
\label{eq:jsimplify}
\end{equation}
where
\begin{equation}
f(\mathbf{w})= -\frac{1}{L}\underset{x\in\mathcal{D}}{\sum}N(x\mathbf{w})\cdot \ln{P(x|\mathbf{w})}+a\cdot \frac{\ln{L}}{L} \cdot s(\mathbf{w})
\end{equation}
$f(\mathbf{w})$ is the cost of end node $\mathbf{w}$ on $\mathcal{T}$. 

We minimize (\ref{eq:jsimplify}) recursively by dividing the original problem into sub-problems.
The sub-problem is to minimize the end node cost $F(\mathcal{T}_{\mathbf{u}}^{0})$ of a sub-tree $\mathcal{T}_{\mathbf{u}}^{0}$ rooted at node $\mathbf{u}$.
Specifically, we define a recurisve function $J(\mathcal{T}_{\mathbf{u}}^{0})$
that minimizes $F(\mathcal{T}_{\mathbf{u}}^{0})$ as follows:
\begin{equation}
J(\mathcal{T}_\mathbf{u}^{0})= \min \{ J(\mathcal{T}_{\mathbf{u}l}^{0})+  J(\mathcal{T}_{\mathbf{u}s}^{0})+  J(\mathcal{T}_{\mathbf{u}r}^{0}),f(\mathbf{u}) \}
\label{eq:jprime_recursive}
\end{equation}
In words, (\ref{eq:jprime_recursive}) states that we can either treat node $\mathbf{u}$ as an end node and compute its cost $f(\mathbf{u})$ (and as a result eliminating all nodes in the sub-tree below), or treat it as an intermediate node and recurse.
The complexity of (\ref{eq:jprime_recursive})---the total number of recursions---with initial tree $\mathcal{T}^0$ as argument is proportional to the number of nodes, hence $O(K)$.

\subsubsection{Analysis of the Optimization Algorithm}
\label{subsubsec:anlysis_optimization}

To better understand the relationship between the likelihood and the depth of the context tree, for an intermediate node $\mathbf{u}$ with three end-node children on tree $\mathcal{T}$, we examine the change in likelihood if the three end-node children are pruned.
%\red{what does this mean?}
%\blue{For an intermediate node $\mathbf{u}$ with three \textit{end node} children on a context tree $\mathcal{T}$, I want to calculate the difference of the likelihood of $\mathcal{T}$ when the three end node children are pruned compared to that the three children are kept on the tree.} 
Following (\ref{eq:jmap_ln}), the change in likelihood is calculated as follows:
%The likelihood difference of the children $\mathbf{u}l$, $\mathbf{u}s$, $\mathbf{u}s$ and node $\mathbf{u}$ is calculated as follows:

\begin{equation}
\begin{split}
&-\frac{1}{L}\underset{v\in\mathcal{D}}{\sum}\underset{x\in\mathcal{D}}{\sum}N(x\mathbf{u}v)\ln{P(x|\mathbf{u}v)} + \frac{1}{L}\underset{x\in\mathcal{D}}{\sum}N(x\mathbf{u})\ln{P(x|\mathbf{u})} \\
&=- \frac{1}{L}\underset{v\in\mathcal{D}}{\sum}N(\mathbf{u}v)\underset{x\in\mathcal{D}}{\sum}P(x|\mathbf{u}v)\ln{\frac{P(x|\mathbf{u}v)}{P(x|\mathbf{u})}}\\
&= -\frac{1}{L}\underset{v\in\mathcal{D}}{\sum}N(\mathbf{u}v)D_{KL}(P(\cdot|\mathbf{u}v)||P(\cdot|\mathbf{u}))
\end{split}
\label{eq:infor_gain}
\end{equation}
where $D_{KL}(P(\cdot|\mathbf{u}v)||P(\cdot|\mathbf{u}))$ is the \textit{Kullback-Leibler divergence} (KLD) \cite{kullback1987letter} of $P(\cdot|\mathbf{u})$ from $P(\cdot|\mathbf{u}v)$, which is non-negative. 
As discussed previously, the average negative log of the likelihood can be regarded as the average information. 
Hence this difference is the average information gain when the children of node $\mathbf{u}$ are pruned. 
In general, we can conclude that the log likelihood term becomes smaller when the tree grows deeper. 
On the other hand, the log prior becomes larger when the tree grows deeper. 
Our objective is thus to find a properly sized context tree that balances these two terms.

\subsection{Adaptive Arithmetic Coding}
\label{subsec:arithmetic_coding}

For each symbol ${x}_i$ in the DCC strings of the target image, we first find the matched context $\mathbf{w}$ of $x_i$, \textit{i.e.}, $\mathbf{w}=\mathbf{x}^{i-1}_{i-|\mathbf{w}|}$, and get the corresponding conditional probability distribution $P(x_i|\mathbf{w})$ from the resulting optimal context tree $\mathcal{T}^{\ast}$. 
%\red{no, u don't compute it again. u have already computed the optimal tree; all the stats are already known. u just retrieve the corresponding computed conditional probability distribution, right?} 
%\blue{Yes, I have rewritten it.} 
Then, the probability distribution is inputted into an adaptive arithmetic coder \cite{DCC1st1991} to encode $x_i$. The length of the DCC string is also losslessly coded using fixed length binary coding.

%% file: lossy.tex
We now discuss lossy contour coding. 
Specifically, we approximate each DCC string in the test image by minimizing an RD cost. The rate of encoding an approximated DCC string is computed using a context tree $\mathcal{T}^{\ast}$, constructed from training data as described previously. 
Two distortion metrics, \textit{sum squared distance distortion} (SSDD) and \textit{maximum absolute distance distortion} (MADD), are introduced for different applications. 
%The approximated DCC strings are coded using arithmetic coding given the same $\mathcal{T}^{\ast}$ used for computing the rate during the minimization process.
We first discuss the two distortion metrics. 
Then we discuss the algorithms of approximating the DCC strings using one of the two metrics in order.

\subsection{Distortion Definition}
\label{ssec:distortion_definition}

When approximating a contour, the chosen distortion metric should be application-specific.
If the intended application is image / video compression, where coded contours are used as side information to facilitate transform coding \cite{hu15} or motion prediction of arbitrarily shaped blocks \cite{daribo14}, then a metric measuring the aggregate distortion between the original and approximated contours would be appropriate. 
SSDD would be a suitable metric in this case.

Denote by $\mathbf{x}$ and $\hat{\mathbf{x}}$ the original and approximated DCC strings respectively, and by $N$ and $\hat{N}$ the lengths of $\mathbf{x}$ and $\hat{\mathbf{x}}$. 
To describe a DCC string geometrically on a 2D grid, we first map $\mathbf{x}$ to a \textit{contour segment}, composed of contiguous vertical or horizontal edges.
Fig.\;\ref{fig:contour_definition}(a) shows a contour segment and the corresponding DCC string.
Specifically, given a default starting point $(0,0)$ on the 2D grid, we determine the $i$-th edge relative to the $(i-1)$-th edge using the $i$-th symbol $x_i$.
Denote by $\mathbf{p}_{\mathbf{x}}(i)$ the 2D coordinate of the endpoint of the $i$-th edge and $a_{\mathbf{x}}(i) \in \mathcal{A}=\{\texttt{N},\texttt{E},\texttt{S},\texttt{W}\}$ the absolute direction of the $i$-th edge. 
%\textit{i.e.}, \texttt{North}, \texttt{East}, \texttt{South} and \texttt{West}.
The $i$-th edge is uniquely determined by the coordinate $\mathbf{p}_{\mathbf{x}}(i)$ and the absolute direction $a_{\mathbf{x}}(i)$ alone.

Denote by $d(\mathbf{p}_{\hat{\mathbf{x}}}(j), \mathbf{x})$ the distortion of the $j$-th approximated symbol $\hat{x}_j$ with respect to the original DCC string $\mathbf{x}$.
$d(\mathbf{p}_{\hat{\mathbf{x}}}(j), \mathbf{x})$ is calculated as the \textit{minimum absolute distance} between coordinate $\mathbf{p}_{\hat{\mathbf{x}}}(j)$ of the $j$-th edge and the segment derived from $\mathbf{x}$ on the 2D grid:
\begin{equation}
d(\mathbf{p}_{\hat{\mathbf{x}}}(j), \mathbf{x})=\underset{1\leq i \leq N}{\min}
\left| \mathbf{p}_{\hat{\mathbf{x}}}(j)-\mathbf{p}_{\mathbf{x}}(i) \right|
\label{eq:error}
\end{equation}
SSDD $D_{S}(\hat{\mathbf{x}}, \mathbf{x})$ is then calculated as the sum of squared distortions of all approximated symbols:
\begin{equation}
D_{S}(\hat{\mathbf{x}}, \mathbf{x})=\sum\limits_{j=1}^{\hat{N}} d^2(\mathbf{p}_{\hat{\mathbf{x}}}(j), \mathbf{x})
\label{eq:D1}
\end{equation}
%\red{do we need to square?}
%\blue{Yes, I think using square would be better. It is also similar to the definition of distortion for depth compression.}

Another distortion metric is MADD, which measures the \textit{maximum} distortion between the original and approximated contours.
MADD is suitable for applications where perceptual quality is evaluated \cite{neuhoff1985rate,lai2010arbitrary,zhu2014adaptive}.
Consider for example a long contour with all edges shifted to the left by one pixel. The contour shift should incur a small perceptual penalty rather than the sum of all individual edge shifts, and so MADD is more reasonable than SSDD in this case.
%Another popular distortion measure is the MADD measure \cite{sohel2012sliding,lai2010arbitrary,zhu2014adaptive} which measures the maximum error between the original and approximated contour.
We calculate MADD $D_{M}(\hat{\mathbf{x}}, \mathbf{x})$ as the maximum distortion of all the approximated symbols:
\begin{equation}
D_{M}(\hat{\mathbf{x}}, \mathbf{x})=\underset{1\leq j \leq \hat{N}}{\max} d(\mathbf{p}_{\hat{\mathbf{x}}}(j), \mathbf{x})
\label{eq:D2}
\end{equation}

\subsection{SSDD based Contour Coding}
\label{ssec:ssdd_contour_coding}

To approximate contour $\mathbf{x}$ using SSDD (\ref{eq:D1}) as the distortion metric, we first write the RD cost as follows:
\begin{equation}
\begin{array}{rl}
\underset{\hat{\mathbf{x}}}{\min} & D_S(\hat{\mathbf{x}}, \mathbf{x}) + \lambda R(\hat{\mathbf{x}})
\end{array}
\label{eq:original_rdo}
\end{equation}
where $R(\hat{\mathbf{x}})$ denotes the overhead to encode DCC string $\hat{\mathbf{x}}$ and $\lambda$ is the Lagrange multiplier.
$R(\hat{\mathbf{x}})$ is approximated as the total information of the symbols in $\hat{\mathbf{x}}$,
\begin{equation}
R(\hat{\mathbf{x}}) = -\sum\limits^{\hat{N}}_{j=1} \log_2 P(\hat{{x}}_j|\hat{\mathbf{x}}^{j-1}_1)
\end{equation}
Given a context tree $\mathcal{T}^{\ast}$, a context $\mathbf{w}$ is selected for each $\hat{x}_j$ to calculate $P(\hat{{x}}_j|\hat{\mathbf{x}}^{j-1}_1)$, where $\mathbf{w}$ is a prefix of $\hat{\mathbf{x}}^{j-1}_1$. Specifically, $P(\hat{{x}}_j|\hat{\mathbf{x}}^{j-1}_1)=P(\hat{x}_j|\mathbf{w})$, calculated as $\frac{N(\hat{x}_j\mathbf{w})}{N(\mathbf{w})}$.

Using the definitions of $D_S(\hat{\mathbf{x}}, \mathbf{x})$ and $R(\hat{\mathbf{x}})$, the objective is written as:
\begin{equation}
F_S(\hat{\mathbf{x}}, \mathbf{x})=\sum\limits_{j=1}^{\hat{N}} d^2(\mathbf{p}_{\hat{\mathbf{x}}}(j), \mathbf{x}) - \lambda \sum\limits^{\hat{N}}_{j=1} \log_2 P(\hat{{x}}_j|\hat{\mathbf{x}}^{j-1}_1)
\label{eq:ssdd_rdo}
\end{equation}

%To keep the connectivity of a closed segment determined by the DCC string on the 2D grid, 
For simplicity, we assume that the 2D coordinates of the first and last edges of the approximated $\hat{\mathbf{x}}$ are the same as those of the original $\mathbf{x}$, \textit{i.e.}, $\mathbf{p}_{\hat{\mathbf{x}}}(1)=\mathbf{p}_{\mathbf{x}}(1)$ and $\mathbf{p}_{\hat{\mathbf{x}}}(\hat{N})=\mathbf{p}_{\mathbf{x}}(N)$.

\subsubsection{Dynamic Programming Algorithm}
\label{sssec:optimization_dynamic_programming}

%Since both distortion and rate in (\ref{eq:ssdd_rdo}) are summations of all symbols in $\hat{\mathbf{x}}$, 
We now describe an efficient DP algorithm to find the optimal approximated contour $\hat{\mathbf{x}}$, minimizing (\ref{eq:ssdd_rdo}).
We first rewrite (\ref{eq:ssdd_rdo}) as:
\begin{equation}
F_S(\hat{\mathbf{x}}, \mathbf{x})=\sum\limits_{j=1}^{\hat{N}} f(\mathbf{p}_{\hat{\mathbf{x}}}(j),\hat{\mathbf{x}}^j_1)
\label{eq:ssdd_rdo_re}
\end{equation}
where 
\begin{equation}
f(\mathbf{p}_{\hat{\mathbf{x}}}(j),\hat{\mathbf{x}}^j_1) =  d^2(\mathbf{p}_{\hat{\mathbf{x}}}(j), \mathbf{x}) - \lambda  \log_2 P(\hat{{x}}_j|\hat{\mathbf{x}}^{j-1}_1)
\label{eq:rd_oneSymbol}
\end{equation}
$f(\mathbf{p}_{\hat{\mathbf{x}}}(j),\hat{\mathbf{x}}^j_1)$ is the RD cost of symbol $\hat{{x}}_j$.
Since $\mathbf{x}$ is fixed, we omit it in the parameters of $f(\,)$ to simplify notations.

Denote by $C_j(\hat{\mathbf{x}}_{j-D}^{j-1},\mathbf{p}_{\hat{\mathbf{x}}}(j-1),a_{\hat{\mathbf{x}}}(j-1))$ the minimum aggregate RD cost from $\hat{x}_j$ to last edge $\hat{x}_{\hat{N}}$ given that the $D$ previous symbols (called \textit{history} in the sequel) are $\hat{\mathbf{x}}_{j-D}^{j-1}$, and the previous edge has coordinate $\mathbf{p}_{\hat{\mathbf{x}}}(j-1)$ and absolute direction $a_{\hat{\mathbf{x}}}(j-1)$. 
%We can write $C_j()$ as:
%\begin{equation}
%C_j(\hat{\mathbf{x}}_{j-D}^{j-1},\mathbf{p}_{\hat{\mathbf{x}}}(j-1),a_{\hat{\mathbf{x}}}(j-1)) = \underset{\hat{{x}}_k \in \mathcal{D}}{\min} \left\{ \sum\limits_{k=j}^{\hat{N}} f(\mathbf{p}_{\hat{\mathbf{x}}}(k),\hat{\mathbf{x}}^{k}_{k-D}) \right\}
%\label{eq:definition_c}
%\end{equation}
$D$ is the maximum depth of $\mathcal{T}^{\ast}$. 
The code rate of $\hat{x}_j$ depends on no more than its $D$ previous symbols in history $\hat{\mathbf{x}}_{j-D}^{j-1}$.

We can calculate $C_j(\hat{\mathbf{x}}_{j-D}^{j-1},\mathbf{p}_{\hat{\mathbf{x}}}(j-1),a_{\hat{\mathbf{x}}}(j-1))$ recursively as follows:
\begin{equation}
\begin{split}
&C_j(\hat{\mathbf{x}}_{j-D}^{j-1},\mathbf{p}_{\hat{\mathbf{x}}}(j-1),a_{\hat{\mathbf{x}}}(j-1)) =\\
&\underset{\hat{x}_j \in \mathcal{D}}{\min} 
\begin{cases}
f(\mathbf{p}_{\hat{\mathbf{x}}}(j),\hat{\mathbf{x}}_{j-D}^{j}), & \mathbf{p}_{\hat{\mathbf{x}}}(j)=\mathbf{p}_{\mathbf{x}}(N) \\
\begin{split}
&f(\mathbf{p}_{\hat{\mathbf{x}}}(j),\hat{\mathbf{x}}_{j-D}^{j}) \\ & + C_{j+1}(\hat{\mathbf{x}}_{j+1-D}^{j},\mathbf{p}_{\hat{\mathbf{x}}}(j),a_{\hat{\mathbf{x}}}(j)),
\end{split}
&\text{o.w.} 
\end{cases}
\end{split}
\label{eq:recursion_ds_original}
\end{equation}
where $\mathbf{p}_{\hat{\mathbf{x}}}(j)$ and $a_{\hat{\mathbf{x}}}(j)$ are the coordinate and absolute direction of the $j$-th edge derived from chosen symbol $\hat{x}_j$.
In words, (\ref{eq:recursion_ds_original}) chooses the next symbol $\hat{x}_j$ to minimize the sum of a local cost $f(\mathbf{p}_{\hat{\mathbf{x}}}(j),\hat{\mathbf{x}}_{j-D}^{j})$ and a recursive cost $C_{j+1}(\,)$ for the remaining symbols in $\hat{\mathbf{x}}$, given updated history $\hat{\mathbf{x}}_{j+1-D}^{j}$ and the $j$-th edge.
If the coordinate of the next edge matches the coordinate of the last edge of $\mathbf{x}$, \textit{i.e.}, $\mathbf{p}_{\hat{\mathbf{x}}}(j)=\mathbf{p}_{\mathbf{x}}(N)$, we terminate the recursion. 

The absolute direction $a_{\hat{\mathbf{x}}}(j)$ and the coordinate $\mathbf{p}_{\hat{\mathbf{x}}}(j)$ of the next edge $j$ are derived from the chosen next symbol $\hat{x}_j$.
$a_{\hat{\mathbf{x}}}(j)$ is the resulting absolute direction after the previous absolute direction $a_{\hat{\mathbf{x}}}(j-1)$ proceeds in the relative direction specified by $\hat{x}_j$.
Coordinate $\mathbf{p}_{\hat{\mathbf{x}}}(j)$ is then derived from $a_{\hat{\mathbf{x}}}(j)$ relative to the previous coordinate $\mathbf{p}_{\hat{\mathbf{x}}}(j-1)$.
For example, given that the previous absolute direction is \texttt{North}, after turning left, \textit{i.e.}, $\hat{x}_j=\texttt{l}$, the resulting absolute direction is \texttt{West}. 
Then $\mathbf{p}_{\hat{\mathbf{x}}}(j)$ is computed as $\mathbf{p}_{\hat{\mathbf{x}}}(j-1)$ going west by one pixel.
With the updated $\mathbf{p}_{\hat{\mathbf{x}}}(j)$ and $\hat{\mathbf{x}}_{j-D}^{j}$, the local cost $f(\mathbf{p}_{\hat{\mathbf{x}}}(j),\hat{\mathbf{x}}_{j-D}^{j})$ is computed using (\ref{eq:rd_oneSymbol}).

In practice, we restrict the approximated DCC string $\hat{\mathbf{x}}$ to be no longer than the original DCC string $\mathbf{x}$, \textit{i.e.}, $\hat{N} \leq N$, to induce a lower rate.
So if $j > N$, we stop the recursion and return infinity to signal an invalid solution. 

%For example, given the previous absolute direction is \texttt{North} ($a_{\hat{\mathbf{x}}}(j-1)=\texttt{N}$), after turning left ($\hat{x}_j=\texttt{l}$), the resulting absolute direction is \texttt{West} ($a_{\hat{\mathbf{x}}}(j)=\texttt{W}$).
%We first update the absolute direction $a_{\hat{\mathbf{x}}}(j)$ given $\hat{x}_j$ relative to the previous absolute direction $a_{\hat{\mathbf{x}}}(j-1)$.
%\begin{equation}
%When $a_{\hat{\mathbf{x}}}(j)$ is \texttt{North}, \texttt{East}, \texttt{South} or \texttt{West}, the next edge goes up, right, down and left by one pixel on the 2D grid, respectively.
%a_{\hat{\mathbf{x}}}(j) = \begin{cases}
%a_{\hat{\mathbf{x}}}(j-1) \Lsh & \hat{{x}}_j=\texttt{l} \\
%a_{\hat{\mathbf{x}}}(j-1)      & \hat{{x}}_j=\texttt{s} \\
%a_{\hat{\mathbf{x}}}(j-1) \Rsh & \hat{{x}}_j=\texttt{r}
%\end{cases}
%\end{equation}

%$\mathbf{p}_{\hat{\mathbf{x}}}(j)$ is then computed based on computed $a_{\hat{\mathbf{x}}}(j)$:
%\begin{equation}
%\mathbf{p}_{\hat{\mathbf{x}}}(j)=\begin{cases}
%\mathbf{p}_{\hat{\mathbf{x}}}(j-1)+[0,-1] & a_{\hat{\mathbf{x}}}(j)=\texttt{N}\\ 
%\mathbf{p}_{\hat{\mathbf{x}}}(j-1)+[1,0] & a_{\hat{\mathbf{x}}}(j)=\texttt{E}\\ 
%\mathbf{p}_{\hat{\mathbf{x}}}(j-1)+[0,1] & a_{\hat{\mathbf{x}}}(j)=\texttt{S}\\ 
%\mathbf{p}_{\hat{\mathbf{x}}}(j-1)+[-1,0] & a_{\hat{\mathbf{x}}}(j)=\texttt{W}
%\end{cases} 
%\label{eq:p_update}
%\end{equation}

\subsubsection{Complexity Analysis}
\label{subsubsec:complexity_DP}

\begin{figure}[t]

\begin{minipage}[b]{1\linewidth}
  \centering
  \centerline{\includegraphics[width=5 cm]{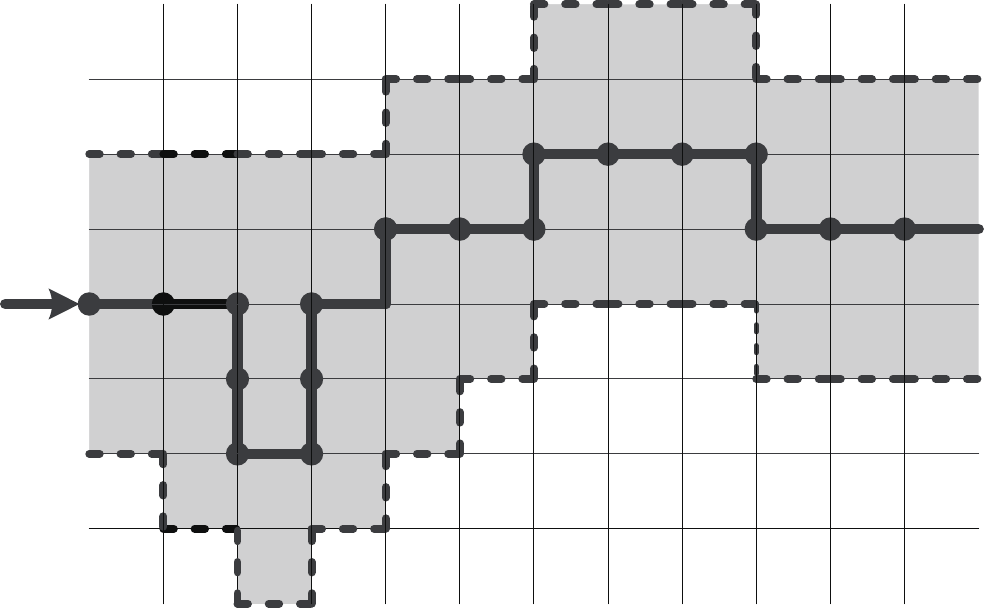}}
  %\vspace{0.1cm}
  \centerline{}
  %\medskip
\end{minipage}

\vspace{-0.2cm}
\caption{An example of possible locations (region in gray) of approximated DCC string defined by $D_{\max}$. In this example, $D_{\max}=2$.}

\label{fig:dmax}
\end{figure}

The complexity of the DP algorithm is upper-bounded by the size of the DP table times the complexity of computing each table entry.
Denote by $Q$ the total number of possible coordinates\footnote{As a computation / quality tradeoff, we can restrict the set of potential coordinates, for example, to be a set of neighborhood grid points within some fixed distance $D_{\max}$ from the original contour segment $\mathbf{x}$; \textit{e.g.}, the set of points in the grey area in Fig.\,\ref{fig:dmax} within $D_{\max}=2$ from original $\mathbf{x}$.} $\mathbf{p}_{\hat{\mathbf{x}}}(j-1)$.
Examining the subscript and three arguments of the recursive function $C_j(\,)$, we see that the DP table size is $N \times 3^D \times Q \times 4$, or $O(N 3^D Q)$.
The complexity of compute one table entry using (\ref{eq:recursion_ds_original}) is $O(1)$. 
Hence the complexity of the algorithm is $O(N 3^D Q)$.

%Since the number of possible coordinate $Q$ is the size of the 2D image grid, which is large. To reduce complexity, we restrict that the error define in (\ref{eq:error}) of one single approximated symbol $\hat{x}_j$ is smaller than a threshold denoted by $D_{\max}$, as illustrated in Fig.\;\ref{fig:dmax}. Hence $Q$ is reduced from the image size to around $(2D_{\max}+1)N$. This restriction will not affect the performance of the dynamic programming algorithm when $D_{\max}$ is large enough. The total complexity thus becomes $O(3^D D_{\max} N^2)$.

\begin{figure}[tb]

\begin{minipage}[b]{1\linewidth}
  \centering
  \centerline{\includegraphics[width=8cm]{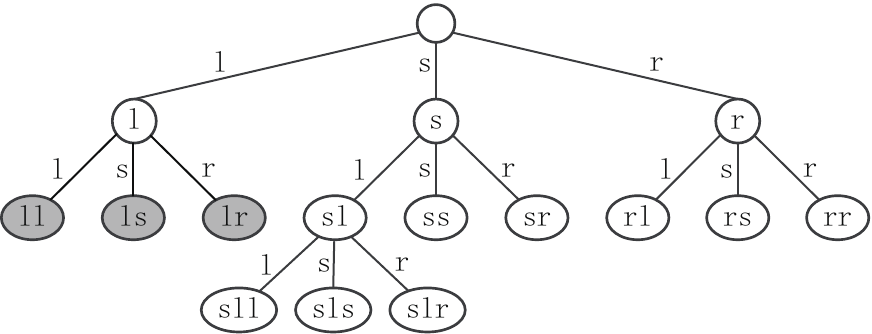}}
%  \vspace{1.5cm}
  \centerline{}\medskip
\end{minipage}

\vspace{-0.3cm}
\caption{An example of \textit{total suffix tree} (TST) derived from the context tree in Fig.\;\ref{fig:MCT_statespace}. End nodes in gray are added nodes based on the context tree. All the end nodes construct a TST: $\mathcal{T}_s^{\ast}=\{\texttt{ll},\texttt{ls},\texttt{lr},\texttt{sll},\texttt{sls},\texttt{slr},\texttt{ss},\texttt{sr},\texttt{rl},\texttt{rs},\texttt{rr}\}$.}
\label{fig:suffix_tree}
\end{figure}

\subsubsection{Total Suffix Tree (TST)}
\label{subsubsec:suffix_tree}

When the training data is large, $D$ is also large, resulting in a very large DP table size due to the exponential term $3^D$.
In (\ref{eq:recursion_ds_original}) when calculating local cost $f(\mathbf{p}_{\hat{\mathbf{x}}}(j),\hat{\mathbf{x}}_{j-D}^{j})$, actually the context required to compute rate is $\mathbf{w}=\hat{\mathbf{x}}_{j-|\mathbf{w}|}^{j-1}$, where the context length $|\mathbf{w}|$ is typically smaller than $D$ because the context tree $\mathcal{T}^*$ of maximum depth $D$ is variable-length.
Thus, if we can, at appropriate recursive calls, reduce the history from $\hat{\mathbf{x}}^{j}_{j+1-D}$ of length $D$ to $\hat{\mathbf{x}}^{j}_{j+1-k}$ of length $k$, $k < D$, for recursive call to $C_{j+1}()$ in (\ref{eq:recursion_ds_original}), then we can reduce the DP table size and in turn the computation complexity of the DP algorithm.

The challenge is how to ``remember" just enough previous symbols $\hat{x}_{j}, \hat{x}_{j-1}, \ldots$ during recursion so that the right context $\mathbf{w}$ can still be correctly identified to compute rate at a later recursive call.
The solution to this problem can be described simply.
Let $\mathbf{w}$ be a context (end node) in context tree $\mathcal{T}^*$.
Context $\mathbf{w}$ must be created at some previous recursive call $C_j()$ by concatenating a chosen $j$-th symbol $\hat{x}_j = w_{|\mathbf{w}|}$ with suffix $\mathbf{w}^{|\mathbf{w}|-1}_1$ of context $\mathbf{w}$.
It implies that the recursion in (\ref{eq:recursion_ds_original}) must remember suffix $\mathbf{w}^{|\mathbf{w}|-1}_1$ for this creation of $\mathbf{w}$ to take place at a later recursion.
To create suffix $\mathbf{w}^{|\mathbf{w}|-1}_1$ at a later recursive call, one must remember its suffix $\mathbf{w}^{|\mathbf{w}|-2}_1$ at an earlier call.
We can thus generalize this observation and state that \textit{a necessary and sufficient condition to preserve all contexts $\mathbf{w}$ in context tree $\mathcal{T}^*$ is to remember all suffixes of $\mathbf{w}$ during the recursion.}

All suffixes of contexts in $\mathcal{T}^*$ can themselves be represented as a tree, which we call a \textit{total suffix tree} (TST), denoted as $\mathcal{T}^*_s$. 
By definition, $\mathcal{T}^*$ is a sub-tree of $\mathcal{T}^*_s$.
Further, TST $\mathcal{T}^*_s$ is also a full tree given $\mathcal{T}^*$ is a full tree; $\mathcal{T}^*_s$ is essentially a union of all sub-trees of $\mathcal{T}^*$, and a sub-tree of a full tree is also a full tree. $\mathcal{T}^*$ has $O(K)$ contexts, each of maximum length $D$. 
Each context can induce $O(D)$ additional end nodes in TST $\mathcal{T}^*_s$. 
Hence TST $\mathcal{T}^*_s$ has $O(K D)$ end-nodes.

Fig.\;\ref{fig:suffix_tree} illustrates one example of TST derived from the context tree shown in Fig.\;\ref{fig:MCT_statespace}.
TST $\mathcal{T}^*_s$ can be used for compact DP table entry indexing during recursion (\ref{eq:recursion_ds_original}) as follows. 
When an updated history $\hat{\mathbf{x}}_{j+1-D}^{j}$ is created from a selection of symbol $\hat{x}_j$, we first truncate $\hat{\mathbf{x}}_{j+1-D}^{j}$ to $\hat{\mathbf{x}}_{j+1-k}^{j}$, where $\hat{\mathbf{x}}_{j+1-k}^{j}$ is the longest matching string in $\mathcal{T}^*_s$ from root node down. 
Because TST $\mathcal{T}^*_s$ is a full tree, the longest matching string always corresponds to an end node. 
The shortened history $\hat{\mathbf{x}}_{j+1-k}^{j}$ is then used as the new argument for the recursive call.
Practically, it means that only DP table entries of arguments $\hat{\mathbf{x}}_{j+1-k}^j$ that are end nodes of TST $\mathcal{T}^*_s$ will be indexed, thus reducing complexity from original $O(N 3^D Q)$ to $O(N K D Q)$, which is now polynomial in $D$.

\subsection{MADD based Contour Coding}

When the distortion metric is MADD, instead of an unconstrained Lagrangian formulation, we formulate instead a distortion-constrained problem as follows:
\begin{equation}
\begin{array}{rl}
\underset{\hat{\mathbf{x}}}{\min} & R(\hat{\mathbf{x}}) \\
s.t. &  D_M(\hat{\mathbf{x}},\mathbf{x}) \leq D_{\max}
\end{array}
\label{eq:dual_rdo}
\end{equation}
where $D_{\max}$ is the maximum distortion permitted. 
Example when $D_{\max} = 2$ is shown in Fig.\;\ref{fig:dmax}.
$D_{\max}$ can be varied to induce different RD tradeoff.

Given the definition of rate and distortion, (\ref{eq:dual_rdo}) can be rewritten as:
\begin{equation}
\begin{array}{rl}
\underset{\hat{\mathbf{x}}}{\min} & -\sum\limits^{\hat{N}}_{j=1} \log_2 P(\hat{{x}}_j|\hat{\mathbf{x}}^{j-1}_1) \\
s.t. &  \underset{1\leq j \leq \hat{N}}{\max} d(\mathbf{p}_{\hat{\mathbf{x}}}(j),\mathbf{x}) \leq D_{\max}
\end{array}
\end{equation}

Similar to the SSDD case, this minimization problem can also be solved by using DP by simplifying the problem to:
\begin{equation}
\begin{array}{rl}
\underset{\hat{\mathbf{x}}}{\min} & \sum\limits^{\hat{N}}_{j=1}  r(\hat{\mathbf{x}}^{j}_1)\\
s.t. &  r(\hat{\mathbf{x}}^{j}_1)=-\log_2 P(\hat{{x}}_j|\hat{\mathbf{x}}^{j-1}_1)\\
 & d(\mathbf{p}_{\hat{\mathbf{x}}}(j), \mathbf{x}) \leq D_{\max}
\end{array}
\label{eq:simplify_dm}
\end{equation}
where $r(\hat{\mathbf{x}}^{j}_1)$ denotes the coding cost of $\hat{{x}}_j$.
The problem becomes finding an approximated DCC string $\hat{\mathbf{x}}$ in the region $\mathcal{R}$ restricted by $D_{\max}$ in order to minimize the total rate. 

Denote by $C_j^{\prime}(\hat{\mathbf{x}}_{j-D}^{j-1},\mathbf{p}_{\hat{\mathbf{x}}}(j-1),a_{\hat{\mathbf{x}}}(j-1))$ the minimum total rate from $\hat{{x}}_j$ to $\hat{{x}}_{\hat{N}}$ given the $D$ previous symbols $\hat{\mathbf{x}}_{j-D}^{j-1}$, and the previous edge has coordinate $\mathbf{p}_{\hat{\mathbf{x}}}(j-1)$ and absolute direction $a_{\hat{\mathbf{x}}}(j-1)$.
We can compute $C_j^{\prime}(\hat{\mathbf{x}}_{j-D}^{j-1},\mathbf{p}_{\hat{\mathbf{x}}}(j-1),a_{\hat{\mathbf{x}}}(j-1))$ recursively as follows:
\begin{equation}
\begin{split}
&C_j^{\prime}(\hat{\mathbf{x}}_{j-D}^{j-1},\mathbf{p}_{\hat{\mathbf{x}}}(j-1),a_{\hat{\mathbf{x}}}(j-1)) =\\
&\underset{\hat{x}_j \in \mathcal{D}}{\min} 
\begin{cases}
r(\hat{\mathbf{x}}_{j-D}^{j}), & \mathbf{p}_{\hat{\mathbf{x}}}(j)=\mathbf{p}_{\mathbf{x}}(N) \\
\begin{split}
&r(\hat{\mathbf{x}}_{j-D}^{j})\\ &+  C_{j+1}^{\prime}(\hat{\mathbf{x}}_{j+1-D}^{j},\mathbf{p}_{\hat{\mathbf{x}}}(j),a_{\hat{\mathbf{x}}}(j)),
\end{split}
& \text{o.w.} 
\end{cases}
\end{split}
\label{eq:recursion_dm}
\end{equation}
where we restrict $\hat{x}_j$ to induce only edges that are within the feasible region $\mathcal{R}$ delimited by $D_{\max}$. 
The recursion is same as (\ref{eq:recursion_ds_original}) except the local cost function. 
Hence the complexity of the DP algorithm here is also same as in the SSDD case.

%% file: start.tex
\begin{figure}[t]

\begin{minipage}[b]{.48\linewidth}
  \centering
  \centerline{\includegraphics[width=4.2cm]{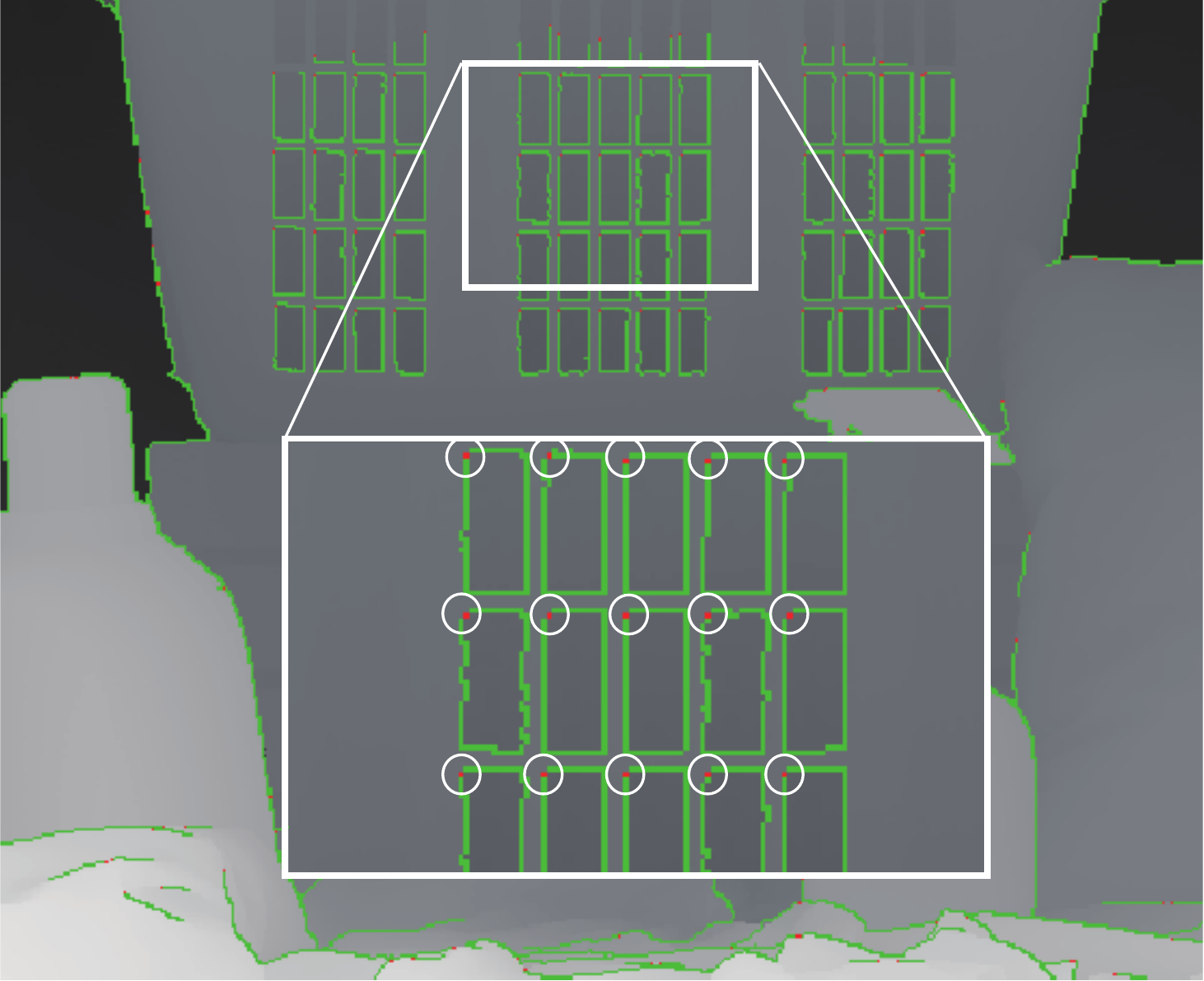}}
%  \vspace{1.5cm}
  \centerline{(a)}\medskip
\end{minipage}
\hfill
\begin{minipage}[b]{0.48\linewidth}
  \centering
  \centerline{\includegraphics[width=4.2cm]{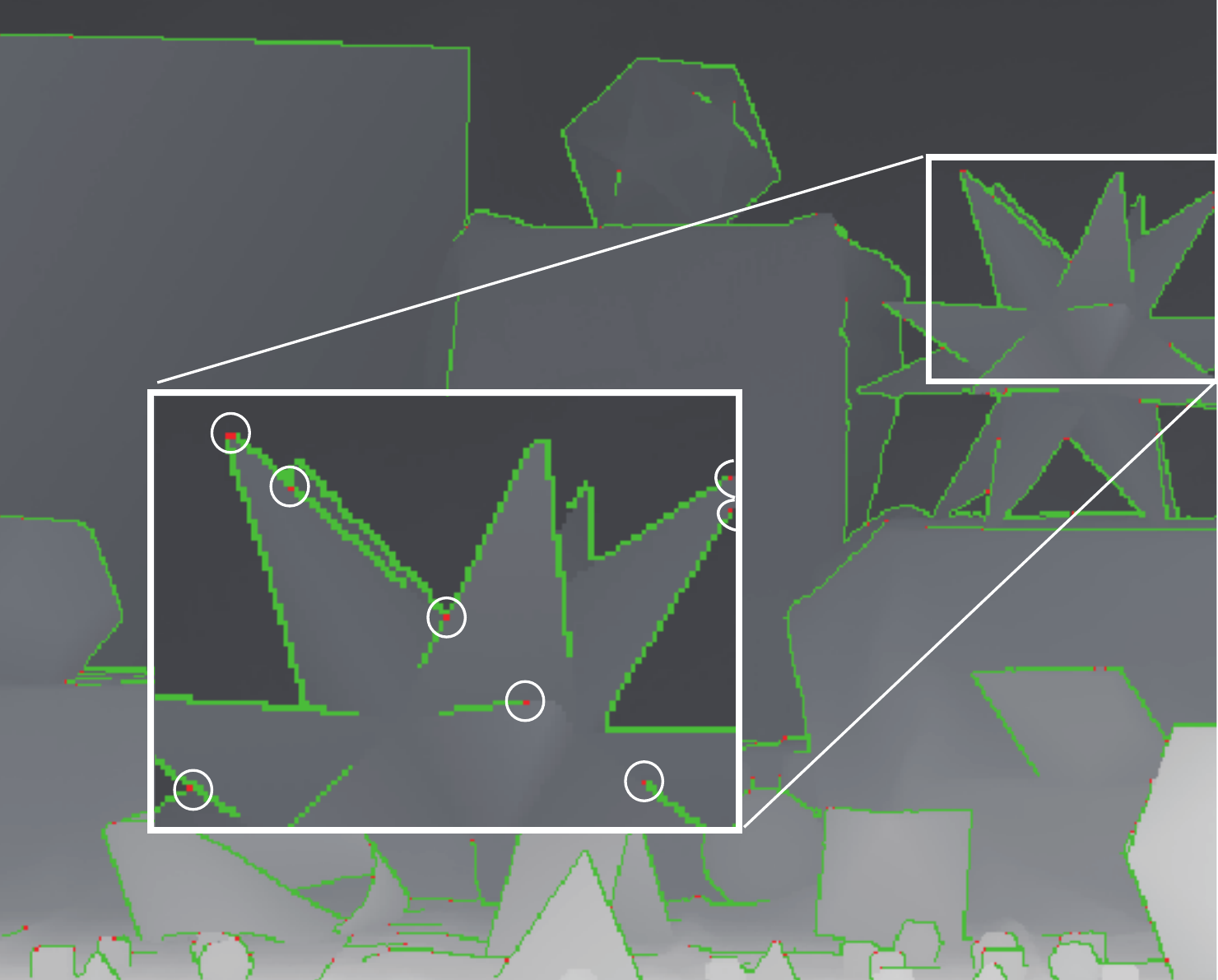}}
%  \vspace{1.5cm}
  \centerline{(b)}\medskip
\end{minipage}

\vspace{-0.2cm}
\caption{Examples of starting points. Green pixels are the pixels along one side of the detected edges and red pixels (inside white circles) are the starting points. (a) \texttt{Laundry}. (b) \texttt{Moebius}.}
\label{fig:startPnt}
\end{figure}

We propose a mixed-Golomb (M-Golomb) algorithm to encode the 2D coordinates of starting points of all contours in the target image. 
%Fixed length binary coding is one choice to losslessly code the coordinates. But in some applications, the target image may contain a lot of DCC strings, fixed length binary coding is expensive to code all the starting points independently.
In general, the starting points are not uniformly distributed on the image.
They tend to cluster around objects or figures that appear in the scene, as illustrated in Fig.\;\ref{fig:startPnt}.
This means that the differences in coordinates of neighboring starting points tend to be small, and Golomb coding \cite{golomb1966} is suitable to encode coordinate differences that are more likely small than large.

A 2D coordinate of a starting point has two components (horizontal and vertical); we use Golomb coding to code only the differences in coordinates in one component. Specifically, we first sort the starting points according to this chosen component in ascending order.
The coordinate differences in neighboring points are then coded using Golomb coding.
The coordinates in the other component are coded using fixed length binary coding.
%We choose the component with larger size to use use Golomb coding, then the component with smaller size will cost fewer bits with fixed length binary coding.

Golomb coding uses a tunable parameter to divide an input value into two parts: the remainder and the quotient. 
The quotient is sent in unary coding, followed by the remainder in truncated binary coding. 
For simplicity, we choose the parameter as $2^k$, where $0 \leq k \leq \ceil{\log_2{W}}$ and $W$ is the maximum input value. 
We examine all possible values of $k$ to find the smallest coding cost of all the starting points in the target image and send this selected parameter as side information.

%% file: results.tex
We evaluate the performance of our lossless and lossy contour coding algorithms in three scenarios. 
%1) lossless contour coding of VOPs in MPEG4 sequences and objects in depth images, 2) lossy contour coding for depth image compressions, and 3) lossy contour coding for multiview silhouettes for 3D reconstruction.
For scenario 1, we first present results for lossless contour coding of VOPs in MEPG4 sequences, then results for lossless contour coding of objects in depth images and starting points coding in order.
For scenario 2, we present visual comparisons of lossy contour coding of objects in depth images, then show the RD performance of lossy depth image coding.
For scenario 3, we present visual comparisons of different silhouette approximating methods, then show the RD performance of lossy multiview silhouette coding for 3D reconstruction.

\subsection{Lossless Contour Coding}
\label{ssec:lossless}

To evaluate the performance of our proposed context tree based lossless contour coding (\texttt{CT-LL}), we used VOPs of four MPEG4 sequences\footnote{\url{ftp://ftp.tnt.uni-hannover.de/pub/MPEG/mpeg4_masks/}}, \texttt{Children}, \texttt{Stefan}, \texttt{Robot} and \texttt{Cyc}, as test data.
The spatial resolution of all four sequences is $352 \times 240$.
10 frames were tested for each sequence.
Note that the contours of the VOPs were already outlined as input to our coding algorithm.
We also coded object contours in four Middleburry depth images\footnote{\url{http://vision.middlebury.edu/stereo/data/}}: \texttt{Moebius} (456$\times $368), \texttt{Cones} (448$\times $368), \texttt{Teddy} (448$\times $368) and \texttt{Laundry} (440$\times $368).
The contours were detected using a gradient-based edge detection scheme in \cite{daribo14}.

To code VOPs in a given video frame of a MPEG-4 sequence, previous two frames of the same sequence were used to train the context tree. 
To code object contours of a Middleburry depth image, we used other Middleburry images to train the context tree. 
%Both the test and the training depth images are obtained using structured light \cite{scharstein2003high}.
We tested two training image sets (\texttt{Train1} and \texttt{Train2}), where each contains four different randomly selected images.
To test the performance of our proposed \texttt{CT-LL} with different training sets, we used also a combined training image set (\texttt{Train1+Train2}), which contains images in both \texttt{Train1} and \texttt{Train2}.
We set the parameter $a = 0.25$ in all the experiments.

We compared \texttt{CT-LL} against four lossless compression schemes: i) the Lempel-Ziv-Welch (\texttt{LZW}) algorithm in \cite{welch1984technique}, an improved version of LZ78; ii) the probability suffix tree (\texttt{PST}) algorithm in \cite{ron1996power}, iii) the prediction by partial matching (\texttt{PPM}) algorithm in \cite{moffat1990implementing}, and iv) the arithmetic edge coding (\texttt{AEC}) scheme in \cite{daribo14}. 
The training datasets for \texttt{PST} and \texttt{PPM} were the same as those for \texttt{CT-LL}.
Our proposed starting points coding scheme (\texttt{M-Golomb}) was compared against fixed length binary coding (\texttt{Binary}).
All the contour compression methods used our proposed \texttt{M-Golomb} for coding the starting points in the experiments.

\begin{table}[htb]
\centering
\caption{Results of Lossless Contour Coding of VOPs in Bits per Symbol}

\begin{center}

\begin{tabular}{c|c|c|c|c|c}
\hline 
\hline
Bits/Symbol & \textbf{LZW} & \textbf{AEC} & \textbf{PST} & \textbf{PPM} & \textbf{CT-LL}\tabularnewline
\hline 
\texttt{Children} & 1.685  & 1.256  & 1.176  & 1.257  & \textbf{1.170}\tabularnewline
\texttt{Stefan} & 1.494  & 1.117  & 0.914  & 0.956  & \textbf{0.894 }\tabularnewline
\texttt{Robot} & 1.808  & 1.414  & 1.341  & 1.318  & \textbf{1.278 }\tabularnewline
\texttt{Cyc} & 1.525  & 1.182  & 0.879  & \textbf{0.823 } & 0.828 \tabularnewline
\hline 
Average & 1.628  & 1.242  & 1.078  & 1.089  & \textbf{1.043 }\tabularnewline
\hline 
\hline
\end{tabular}
\end{center}
\label{tab:lossless_mpeg4_mask}
\end{table}

\begin{table*}[htb] \setlength{\tabcolsep}{2.8pt}
\centering
\caption{Results of Lossless Coding Contours of Objects in Depth Images in Bits per Symbol}

\begin{tabular}{c|c|c|c|c|c|c|c|c|c|c|c}
\hline 
\hline 
\multirow{2}{*}{Bits/Symbol} & \multirow{2}{*}{\texttt{LZW}} & \multirow{2}{*}{\texttt{AEC}} & \multicolumn{3}{c|}{\texttt{Train1}} & \multicolumn{3}{c|}{\texttt{Train2}} & \multicolumn{3}{c}{\texttt{Train1+Train2}}\tabularnewline
\cline{4-12} 
 &  &  & \texttt{PST} & \texttt{PPM} & \texttt{CT-LL} & \texttt{PST} & \texttt{PPM} & \texttt{CT-LL} & \texttt{PST} & \texttt{PPM} & \texttt{CT-LL}\tabularnewline
\hline 
\texttt{Moebius} & 1.912  & 1.409  & 1.378  & 1.272  & \textbf{1.233 } & 1.373  & 1.286  & \textbf{1.240 } & 1.326 & 1.257 & \textbf{1.220}\tabularnewline
\texttt{Cones} & 1.727  & 1.424  & 1.245  & 1.181  & \textbf{1.164 } & 1.282  & 1.232  & \textbf{1.199 } & 1.241 & 1.176 & \textbf{1.154}\tabularnewline
\texttt{Teddy} & 2.014  & 1.519  & 1.555  & 1.493  & \textbf{1.450 } & 1.550  & 1.493  & \textbf{1.434 } & 1.510 & 1.479 & \textbf{1.439}\tabularnewline
\texttt{Laundry} & 1.642  & 1.265  & 1.290  & 1.216  & \textbf{1.189 } & 1.257  & 1.199  & \textbf{1.165 } & 1.255 & 1.195 & \textbf{1.166}\tabularnewline
\hline 
Average & 1.824  & 1.404  & 1.367  & 1.291  & \textbf{1.259 } & 1.366  & 1.303  & \textbf{1.260 } & 1.333 & 1.277 & \textbf{1.245}\tabularnewline
\hline 
\hline 
\end{tabular}

\label{tab:lossless_middleburry_depth}
\end{table*}

\subsubsection{Results for Lossless Contour Coding of VOPs in MPEG4 Sequences} 

Table\;\ref{tab:lossless_mpeg4_mask} shows the compression results in average bits per symbol for VOP lossless coding using different methods.
Compared to the other methods, our proposed \texttt{CT-LL} achieves noticeable bitrate reduction on average.
Specifically, on average we reduce the bitrate by 35.97\% compared to \texttt{LZW}, 16.08\% compared to \texttt{AEC}, 3.26\% compared to \texttt{PST}, and 4.23\% compared to \texttt{PPM}.

In general, the context based methods (\texttt{PST}, \texttt{PPM} and \texttt{CT-LL}) outperform \texttt{LZW} and \texttt{AEC} significantly.
Compared to \texttt{PPM}, the gain of our \texttt{CT-LL} varies depending on different sequences.
In particular, for \texttt{Children} and \texttt{Stefan} with large motion, \texttt{CT-LL} outperforms \texttt{PPM} by 6.92\% and 6.49\%.
While for \texttt{Cyc} with very small motion, \texttt{CT-LL} is worse than \texttt{PPM} marginally by 0.60\%.
Since there is no prior to prune the contexts in \texttt{PPM}, it fails to avoid overfitting for sequences with large motion.
Compared to \texttt{PST}, which can avoid overfitting by setting five application-specific thresholds, \texttt{CT-LL} has steady gain because of the our proposed geometric prior for contour coding.

\begin{table*}[t]

\centering
\caption{Results of Coding Starting Points}

\begin{tabular}{c|c|c|c|c|c|c|c|c}
\hline 
\hline 
\multirow{2}{*}{} & \multirow{2}{*}{Input} & Average Num & \multicolumn{3}{c|}{Bits of Starting Points} & \multicolumn{3}{c}{Total Bits}\tabularnewline
\cline{4-9} 
 &  & of Contours & \texttt{Binary} & \texttt{M-Golomb} & $\triangle$Bits & \texttt{Binary} & \texttt{M-Golomb} & $\triangle$Bits\tabularnewline
\hline 
 & \texttt{Children} & 3.4 & \textbf{57.8} & 58.7 & 1.56\% & \textbf{1684.7} & 1685.6 & 0.05\%\tabularnewline
Mask & \texttt{Stefan} & 1 & \textbf{17} & 21 & 23.53\% & \textbf{517.1} & 521.1 & 0.77\%\tabularnewline
Seqeunces & \texttt{Robot} & 21.3 & 362.1 & \textbf{273.8} & -24.39\% & 2474.1 & \textbf{2385.8} & -3.57\%\tabularnewline
 & \texttt{Cyc} & 21.6 & 367.2 & \textbf{295.5} & -19.53\% & 2959.5 & \textbf{2887.8} & -2.42\%\tabularnewline
\cline{2-9} 
 & Average & 11.8 & 201.0 & \textbf{162.3} & -19.29\% & 1908.9 & \textbf{1870.1} & -2.03\%\tabularnewline
\hline 
 & \texttt{Moebius} & 144 & 2592 & \textbf{1785} & -31.13\% & 8825 & \textbf{8018} & -9.14\%\tabularnewline
Depth & \texttt{Cones} & 82 & 1476 & \textbf{1081} & -26.76\% & 7399 & \textbf{7004} & -5.34\%\tabularnewline
Images & \texttt{Teddy} & 100 & 1800 & \textbf{1289} & -28.39\% & 7894 & \textbf{7467} & -5.41\%\tabularnewline
 & \texttt{Laundry} & 125 & 2250 & \textbf{1563} & -30.53\% & 9254 & \textbf{8567} & -7.42\%\tabularnewline
\cline{2-9} 
 & Average & 112.8 & 2029.5 & \textbf{1429.5} & -29.56\% & 8343.0 & \textbf{7764.0} & -6.94\%\tabularnewline
\hline 
\hline 
\end{tabular}

\label{tab:starting_points}
\end{table*}

\subsubsection{Results for Lossless Contour Coding of Objects in Middlebury Depth Images}

Table\;\ref{tab:lossless_middleburry_depth} shows the compression results for lossless contour coding of objects in depth images.
Our \texttt{CT-LL} outperforms the other four methods for all three training sets.
Specifically, for \texttt{Train1}, the average bit reductions are 30.96\%, 10.35\%, 7.89\% and 2.45\% over \texttt{LZW}, \texttt{AEC}, \texttt{PST} and \texttt{PPM}, respectively;
for \texttt{Train2}, the average bit reductions are 30.93\%, 10.30\%, 7.76\% and 3.31\%;
for \texttt{Train1+Train2}, the average bit reductions are 31.74\%, 11.36\%, 6.62\% and 2.49\%.

Compared to the results of lossless contour coding of VOPs, the performance of \texttt{CT-LL} of lossless contour coding of objects in depth images decreased a bit.
The difference mainly stems from the dissimilarity in statistics between the test and training images in the latter case.
Specifically, for coding VOPs in MPEG4 video sequences, the training images are the previous frames, which are quite similar to the current frame.
While for the test depth images, the training images are randomly selected images, which were obtained using the same setup but from different 3D scenes.
Nonetheless, as shown in Table\;\ref{tab:lossless_middleburry_depth}, we can still achieve bitrate reduction compared to other methods using different training sets.

Comparing the results of \texttt{Train1+Train2} to that of \texttt{Train1} and \texttt{Train2}, we can see that all three context based methods (\texttt{PST}, \texttt{PPM} and \texttt{CT-LL}) benefited from more training data, resulting in lower average bitrates.
With twice the training data, \texttt{PST} and \texttt{PPM} resulted in a larger improvement compared to \texttt{CT-LL}.
This demonstrates that the performance of our proposed \texttt{CT-LL} is stable for large and small data size. 
In other words, unlike other context based methods like \texttt{PST} and \texttt{PPM}, our \texttt{CT-LL} maintains good coding performance even for small data size by avoiding overfitting using our proposed geometric prior.

\subsubsection{Results for Starting Points Coding}

The performance of our proposed starting points coding scheme (\texttt{M-Golomb}) compared to the fixed binary coding (\texttt{Binary}) is shown in Table\;\ref{tab:starting_points}.
We tested both the VOPs in MPEG4 sequences and the object contours in depth images.
Both the number of contours and the bit overhead in coding contours of VOPs are averaged to one frame to better compare to that of the depth images.
For the VOPs which contain only a few contours in one frame, \textit{i.e.}, \texttt{Children} and \texttt{Stefan}, the proposed \texttt{M-Golomb} has little advantage over \texttt{Binary}.
However, we can save on average 29.56\% bits of starting points and 6.94\% bits of total bits for the depth images which contain lots of contours in one image.

\subsection{Depth Image Coding}
\label{ssec:lossy_SSDD}

We implemented our proposed SSDD based lossy contour coding (\texttt{CT-SSDD}) scheme for coding edges as side information in \texttt{MR-GFT} \cite{hu15}, one of the state-of-the-art depth image compression methods.
In \cite{hu15}, the detected edges of the whole image are losslessly coded by employing \texttt{AEC} \cite{daribo14} and transmitted to the decoder for directional intra prediction.
We replace \texttt{AEC} with \texttt{CT-SSDD} to compress four Middleburry depth images which are same as that in section \ref{ssec:lossless}.
\texttt{Train2} are used as the training images.

\begin{figure}[htb]

\begin{minipage}[b]{.48\linewidth}
  \centering
  \centerline{\includegraphics[width=4.5cm]{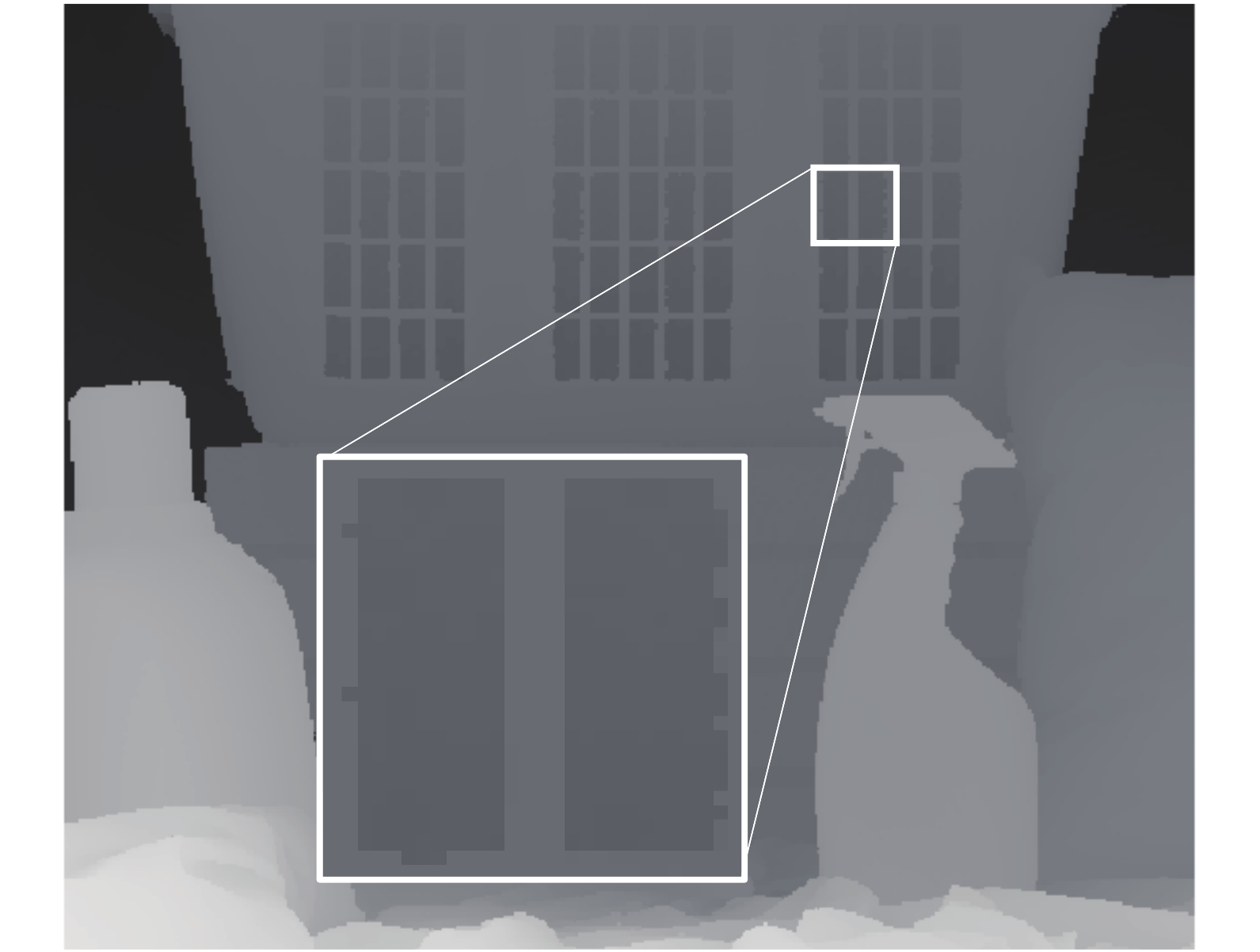}}
%  \vspace{1.5cm}
  \centerline{(a)}\medskip
\end{minipage}
\hfill
\begin{minipage}[b]{0.48\linewidth}
  \centering
  \centerline{\includegraphics[width=4.5cm]{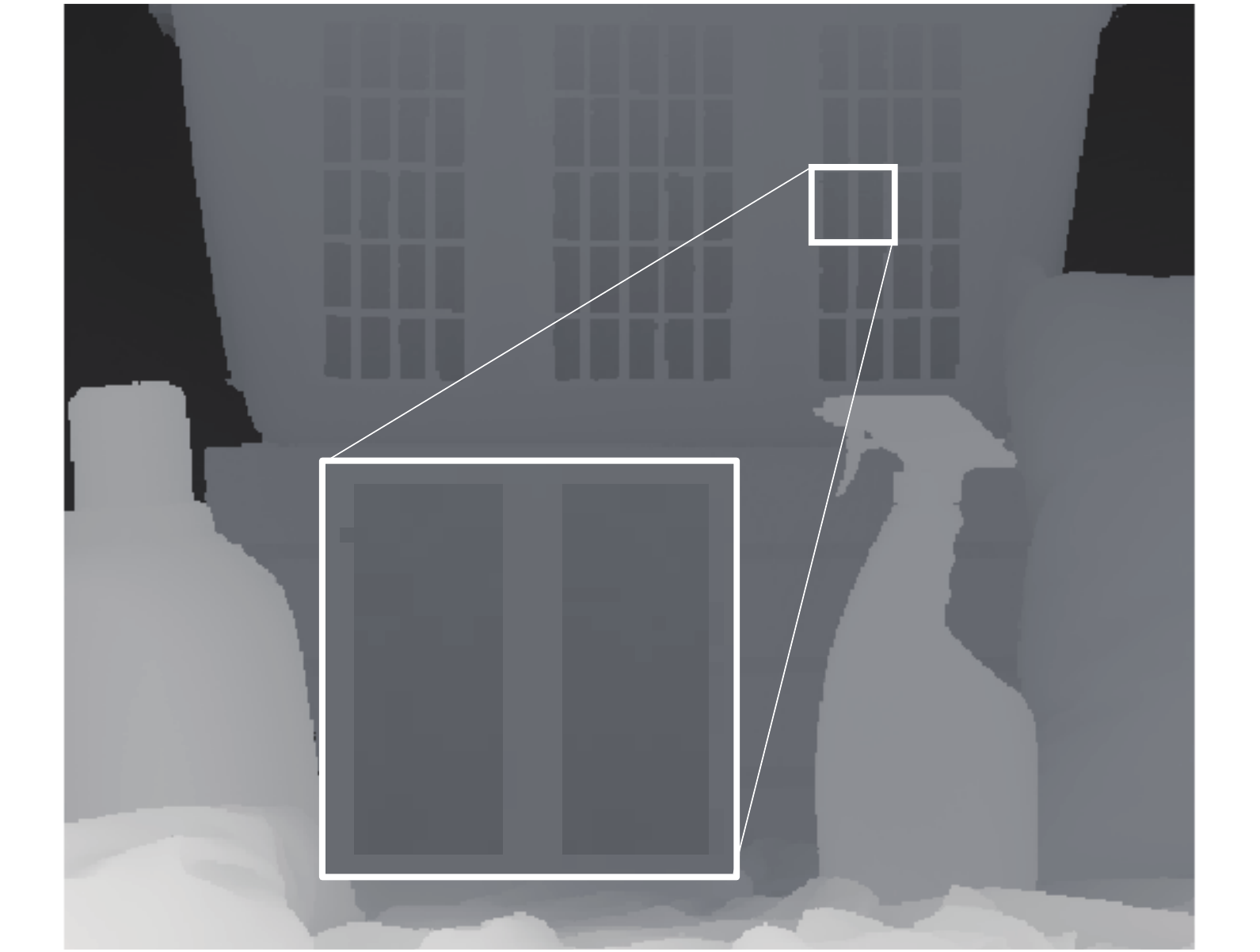}}
%  \vspace{1.5cm}
  \centerline{(b)}\medskip
\end{minipage}
\vfill
\begin{minipage}[b]{.48\linewidth}
  \centering
  \centerline{\includegraphics[width=4.5cm]{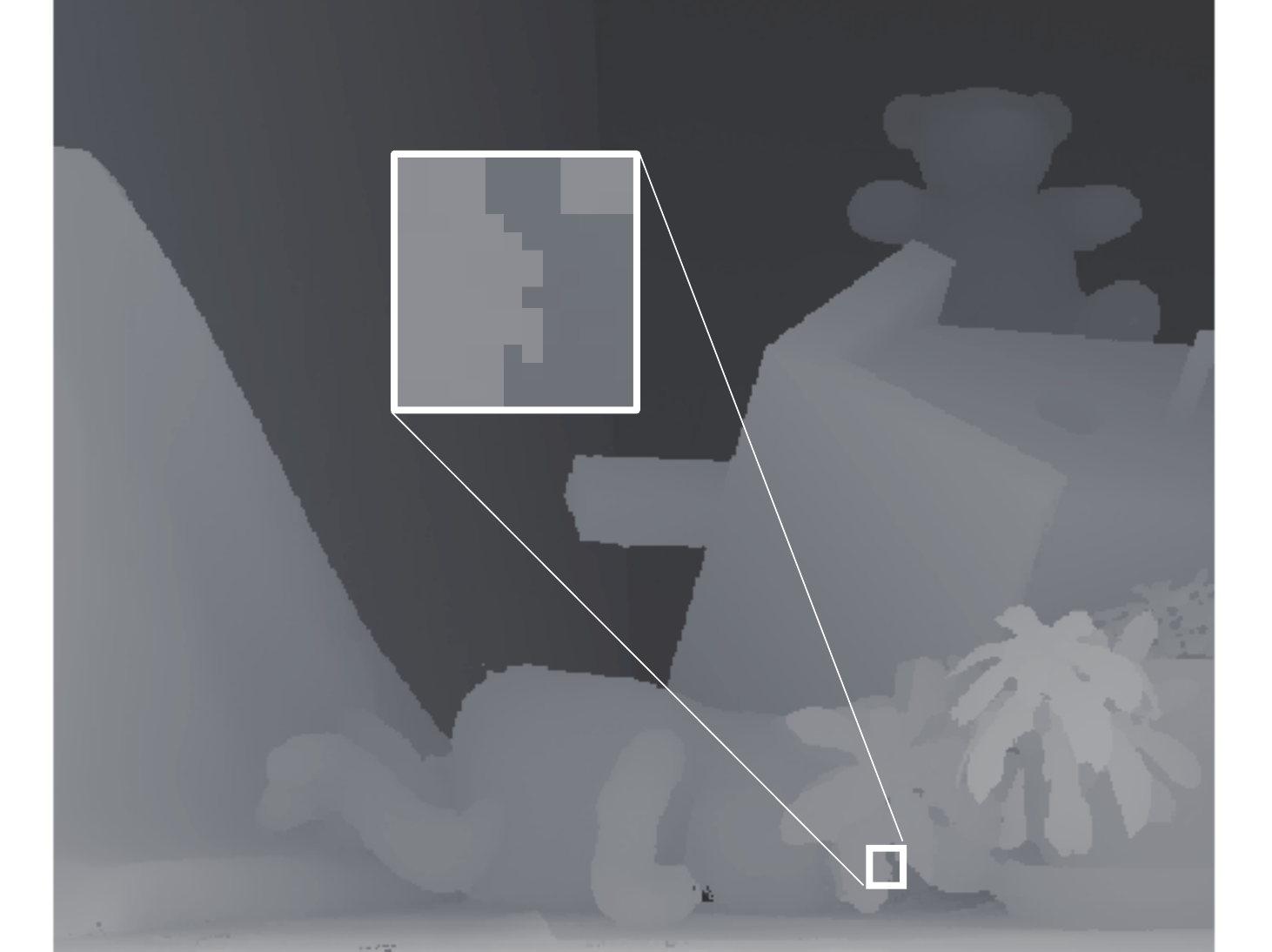}}
%  \vspace{1.5cm}
  \centerline{(c)}\medskip
\end{minipage}
\hfill
\begin{minipage}[b]{0.48\linewidth}
  \centering
  \centerline{\includegraphics[width=4.5cm]{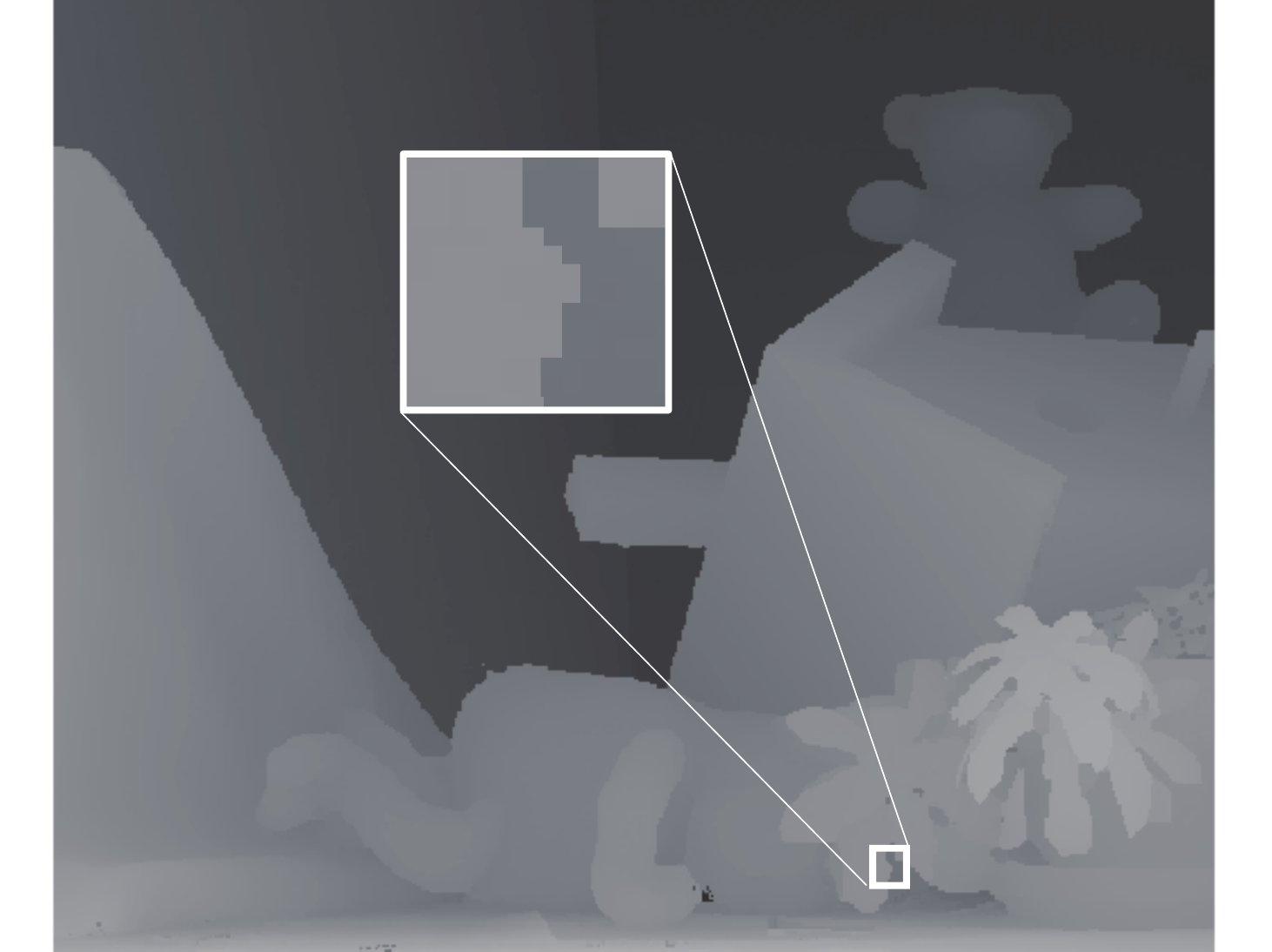}}
%  \vspace{1.5cm}
  \centerline{(d)}\medskip
\end{minipage}

\vspace{-0.3cm}
\caption{Examples of edge approximation and depth image augmentation. (a) Original depth image of \texttt{Laundry}. (b) Augmented depth image of \texttt{Laundry} after edge approximation with $QP=37$. (c) Original depth image of \texttt{Teddy}. (d) Augmented depth image of \texttt{Teddy} after edge approximation with $QP=37$.}
\label{fig:edge_approximation}
\end{figure}

Different from the losslessly edge coding in \texttt{MR-GFT}, we compressed contours lossily using our proposed \texttt{CT-SSDD}.
Changes in the contours can lower the edge coding cost, but can also lead to larger intra prediction residual.
We thus augmented the depth image to match the approximated edges obtained using \texttt{CT-SSDD}, in order to reduce the prediction residual.
Specifically, the pixel values between the original edge and the approximated edge were replaced with the pixel values on the other side of the approximated edge, as shown on Fig.\;\ref{fig:edge_approximation}. 

Since the contours are coded as side information for compressing depth images, the distortion term in \texttt{CT-SSDD} should be related to the distortion of the depth signal.
Distortion of our described depth image coding scheme comes from two sources: i) distortion due to quantization of the transform coefficients, and ii) distortion due to depth pixel augmentation to suit approximated contours, as described previously.
Because the contour coding and the residual coding in \texttt{MR-GFT} are performed in separate steps, it is impractical to consider both distortions simultaneously.

As a straightforward attempt, the distortion term $D_{S}(\hat{\mathbf{x}}, \mathbf{x})$ in (\ref{eq:D1}) is modified and defined as the augmentation distortion, which is calculated as the sum of squared distortion of augmented depth pixels instead of the approximated symbols.
Further, we choose $\lambda=0.85\times(2^\frac{QP-12}{3})$, which is the same $\lambda$ as used in the mode selection part of \texttt{MR-GFT}, where $QP$ is the quantization parameter.
Thus this selection of $\lambda$ provides an appropriate weight between the edge coding and the residual coding.
%Besides, $\lambda$ is positive correlated to $QP$, which is able to adjust the strength of the approximation based on different level of bit rate.
%In other words, the edge can be more approximated with larger $QP$.
%Regarding to the complexity concern, we divide the contour into segments with length 20.

We compare the depth image coding performance of our proposed \texttt{MR-GFT+CT-SSDD} against \texttt{MR-GFT+CT-LL}, \texttt{MR-GFT+PPM}, the original \texttt{MR-GFT} in \cite{hu15} and HEVC Test Model HM-16.0\footnote{\url{https://hevc.hhi.fraunhofer.de/svn/svn_HEVCSoftware/tags/HM-16.0/}} (\texttt{HEVC}), where \texttt{MR-GFT+CT-LL} and \texttt{MR-GFT+PPM} are obtained by replacing the edge coding scheme (\texttt{AEC}) in \texttt{MR-GFT} with the proposed \texttt{CT-LL} and \texttt{PPM}.

\subsubsection{Edge Approximation and Depth Image Augmentation}

Fig.\;\ref{fig:edge_approximation} shows two examples of edge approximation and depth image augmentation.
We see that the changed edges are mostly the irregular edges (e.g., the white region in Fig.\;\ref{fig:edge_approximation}).
This is because these irregular edges along with their contexts have more scattered probability distributions in the trained context tree $\mathcal{T}^{\ast}$, which will consume larger amount of bits after arithmetic coding.
As shown in Fig.\;\ref{fig:edge_approximation}, after approximating the edges, the depth pixels are augmented to match the new edges.

Note that edges with smaller gradient (e.g., mesh grid of basket in Fig.\;\ref{fig:edge_approximation}(b) and flower in Fig.\;\ref{fig:edge_approximation}(d)) are more likely to be approximated than edges with larger gradient (e.g., the boundaries of basket in Fig.\;\ref{fig:edge_approximation}(b) and Teddy bear in Fig.\;\ref{fig:edge_approximation}(d)).
This is because approximation of the edges with larger gradient will result in larger augmentation distortion.

%One example of the approximated edges with edge distortion, which is defined between the original edges and the approximated edges, is shown in Fig. XX. Compared to the result using depth augmentation distortion, more edges are changed with the edge distortion. The reason is that with the same changes of edges, the depth augmentation distortion tends to be much larger than the edge distortion. Under the same bits of coding edges and the same $\lambda$, the edges have smaller chance to be changed using the augmentation distortion. 

\begin{table}[h] \setlength{\tabcolsep}{3pt}
\centering
\caption{Results of Bit Allocation for Coding \texttt{Teddy}}

\begin{tabular}{c|c|c|c|c|c|c|c|c|c}
\hline 
\hline 
\multirow{3}{*}{QP} & \multicolumn{3}{c|}{\texttt{MR-GFT}} & \multicolumn{3}{c|}{\texttt{MR-GFT}} & \multicolumn{3}{c}{\texttt{MR-GFT}}\tabularnewline
 & \multicolumn{3}{c|}{} & \multicolumn{3}{c|}{\texttt{CT-LL}} & \multicolumn{3}{c}{\texttt{CT-SSDD}}\tabularnewline
\cline{2-10} 
 & Edge & Resi & Total & Edge & Resi & Total & Edge & Resi & Total\tabularnewline
\hline 
22 & 6515 & \textbf{12132} & 18647 & 6178 & \textbf{12132} & 18310 & \textbf{6160} & \textbf{12132} & \textbf{18292}\tabularnewline
27 & 6515 & 7857 & 14372 & 6178 & 7857 & 14035 & \textbf{6101} & \textbf{7854} & \textbf{13955}\tabularnewline
32 & 6515 & 6225 & 12740 & 6178 & 6225 & 12403 & \textbf{5827} & \textbf{6021} & \textbf{11848}\tabularnewline
37 & 6515 & 5344 & 11859 & 6178 & 5344 & 11522 & \textbf{5548} & \textbf{5322} & \textbf{10870}\tabularnewline
\hline 
\hline 
\end{tabular}

\label{tab:SSDD_edge_bits}
\end{table}

\begin{figure}[htb]

\begin{minipage}[b]{.48\linewidth}
  \centering
  \centerline{\includegraphics[width=4.5cm]{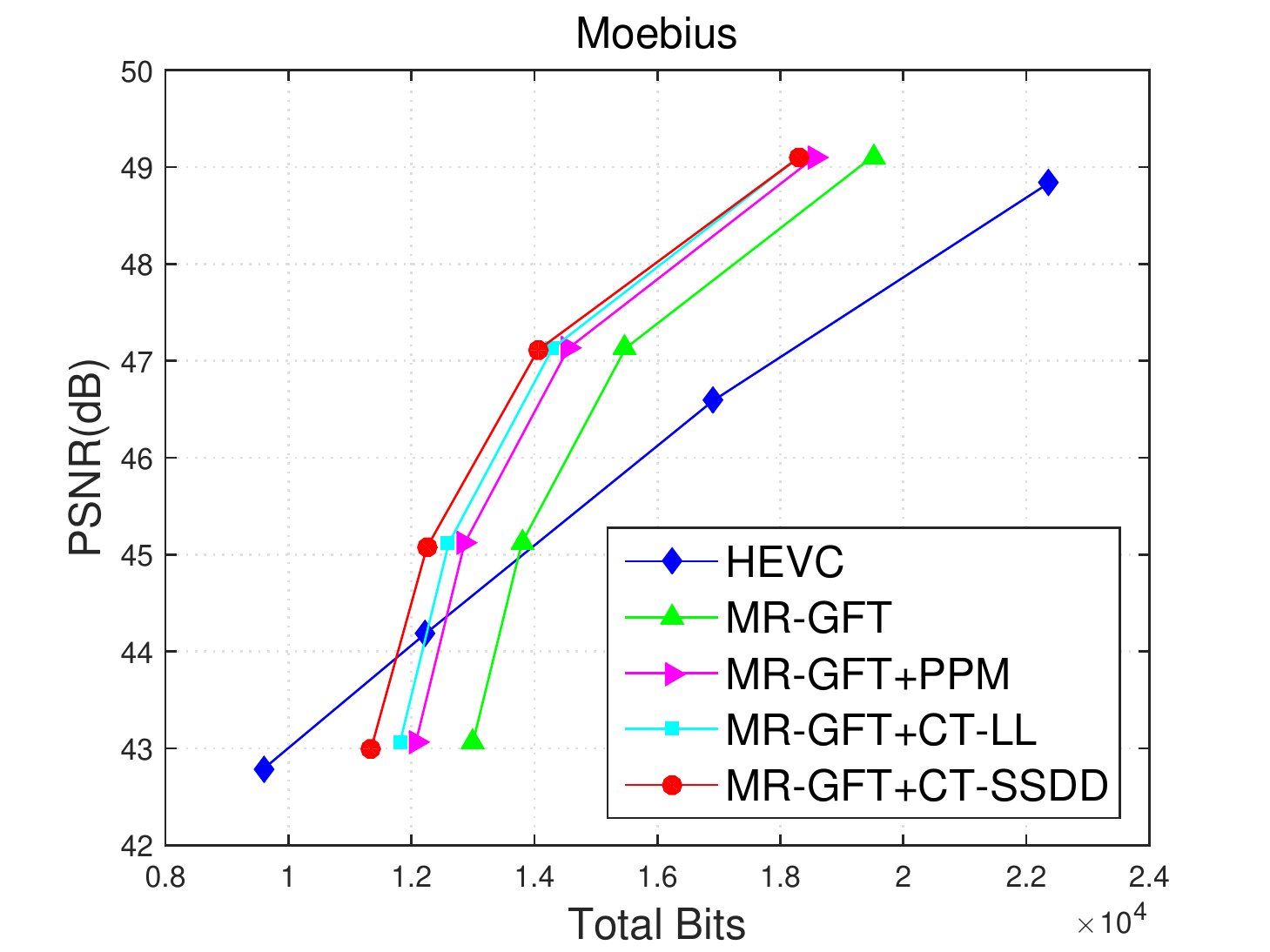}}
%  \vspace{1.5cm}
  \centerline{(a)}\medskip
\end{minipage}
\hfill
\begin{minipage}[b]{0.48\linewidth}
  \centering
  \centerline{\includegraphics[width=4.5cm]{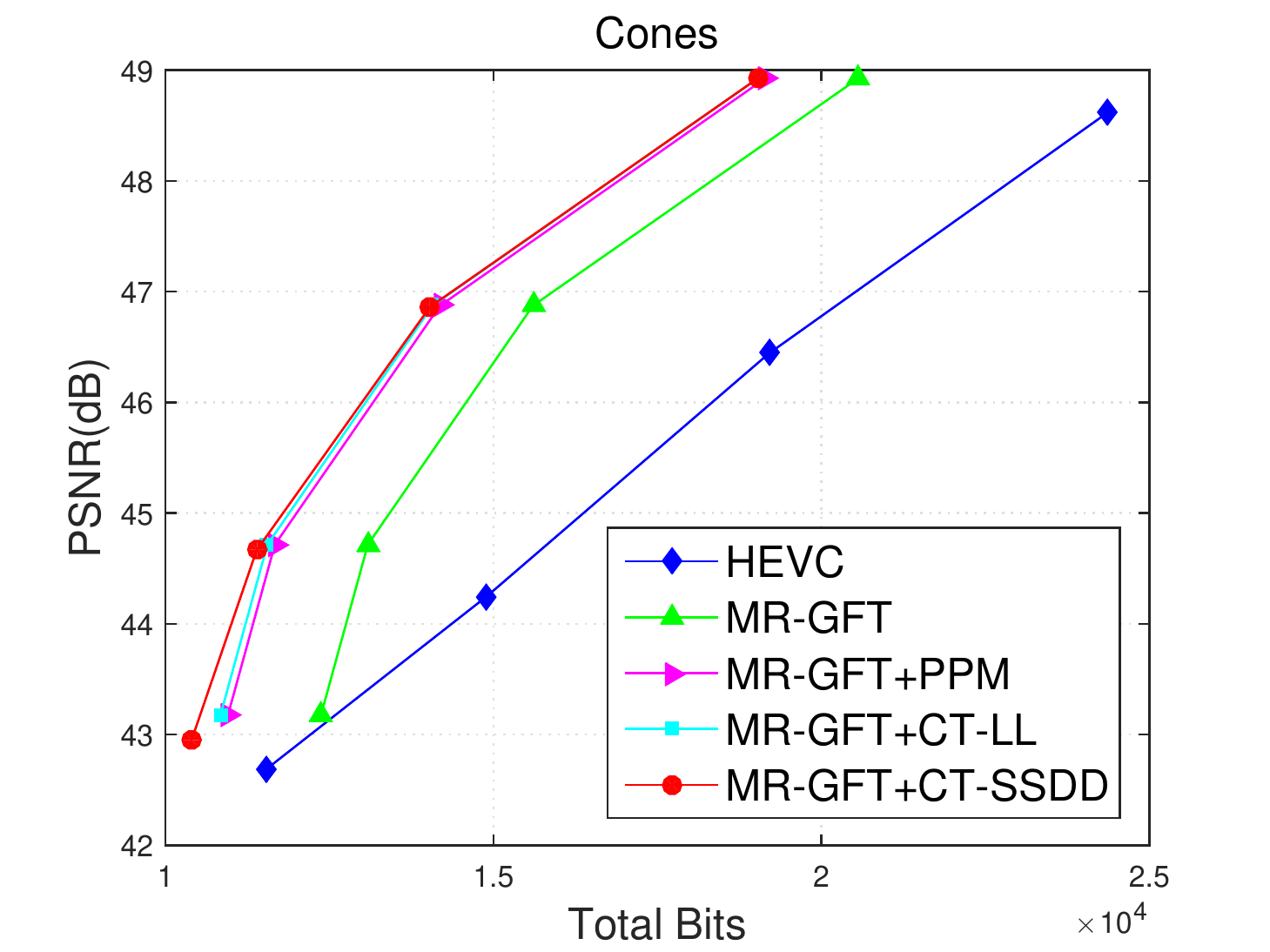}}
%  \vspace{1.5cm}
  \centerline{(b)}\medskip
\end{minipage}
\vfill
\begin{minipage}[b]{.48\linewidth}
  \centering
  \centerline{\includegraphics[width=4.5cm]{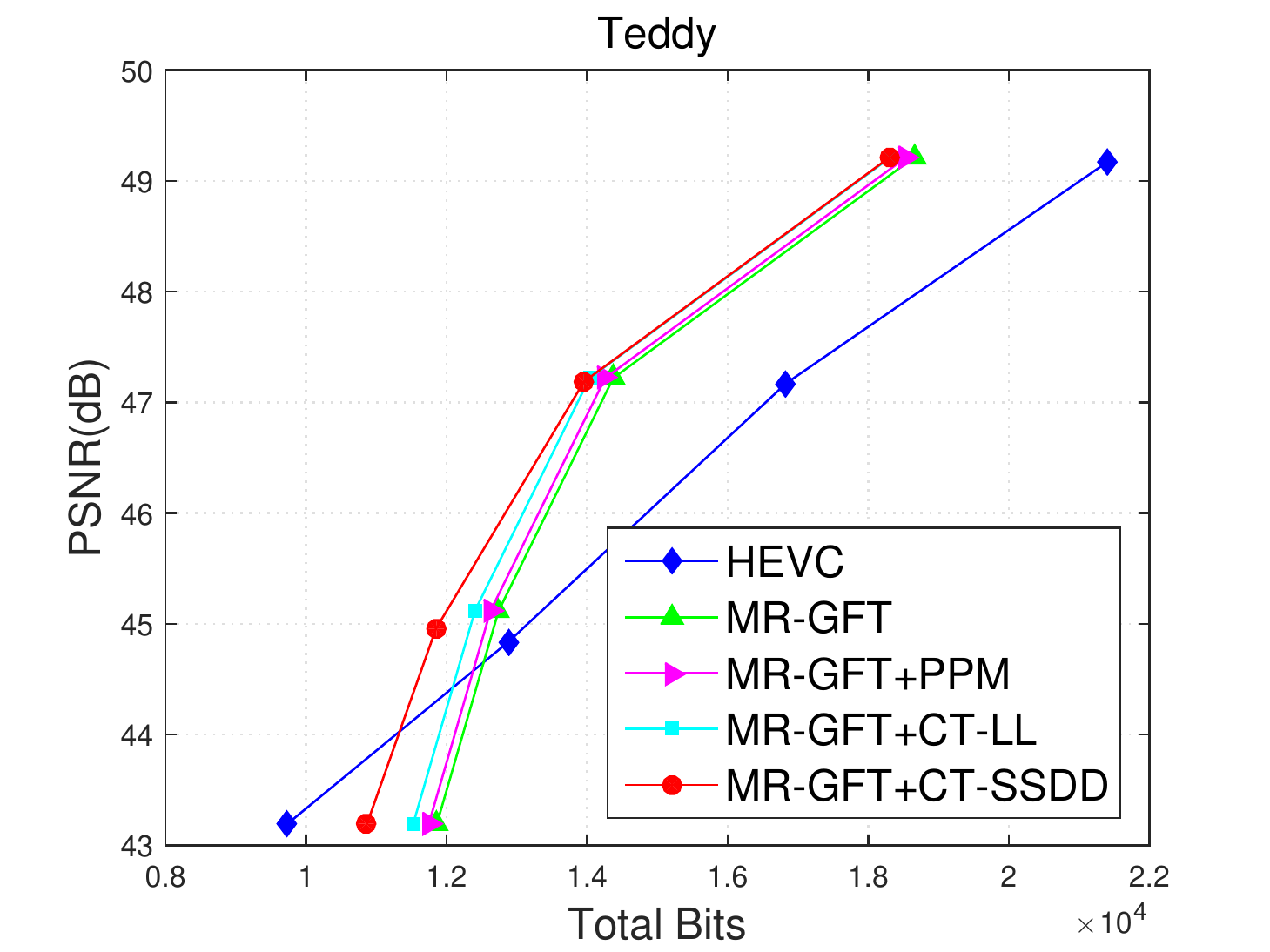}}
%  \vspace{1.5cm}
  \centerline{(c)}\medskip
\end{minipage}
\hfill
\begin{minipage}[b]{0.48\linewidth}
  \centering
  \centerline{\includegraphics[width=4.5cm]{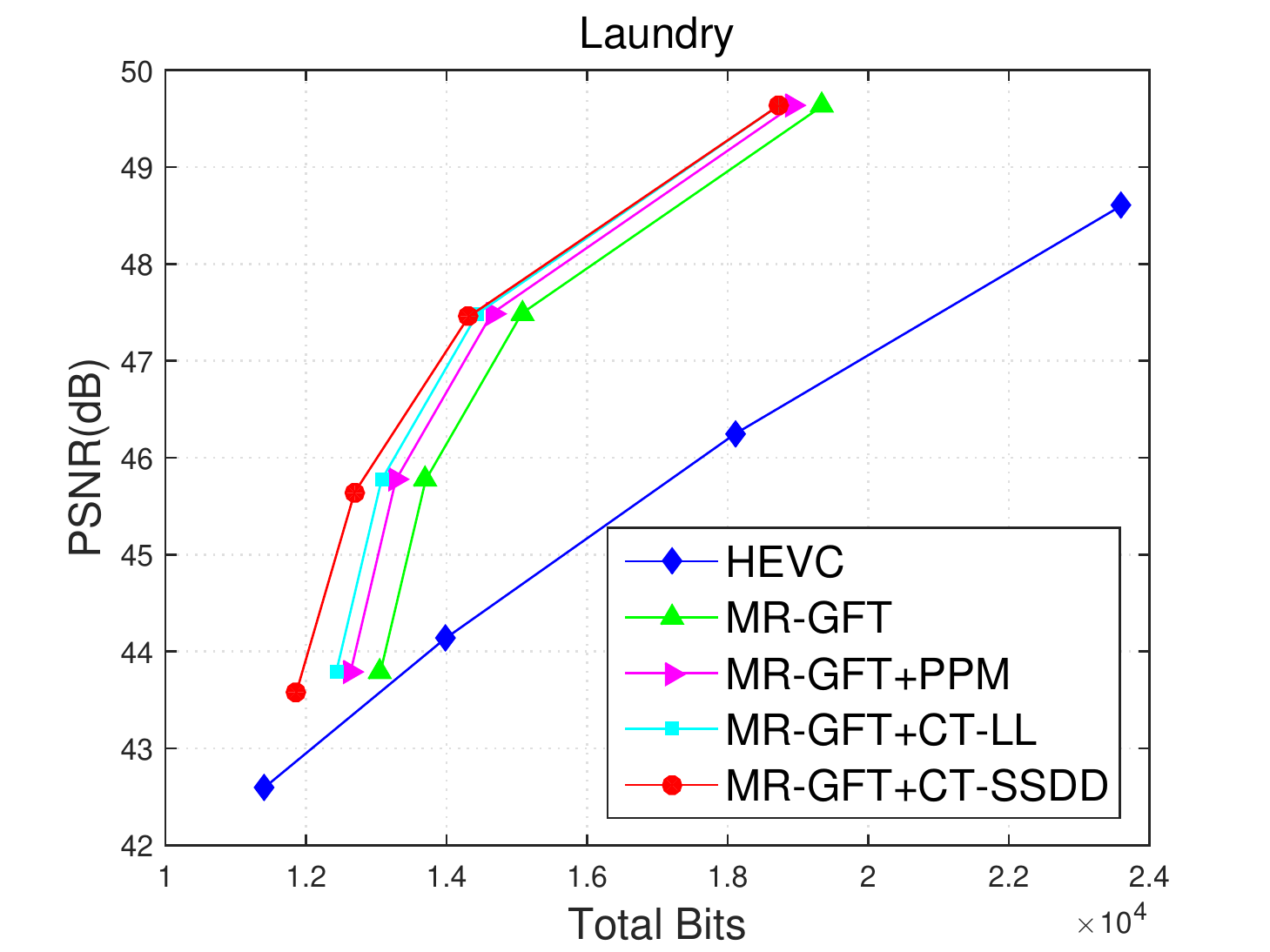}}
%  \vspace{1.5cm}
  \centerline{(d)}\medskip
\end{minipage}

\vspace{-0.3cm}
\caption{RD performance comparison among different compression schemes for depth images.}
\label{fig:RD_Curve}
\end{figure}

\subsubsection{Results in RD Performance}

The RD performance of the proposed \texttt{MR-GFT+CT-SSDD} and \texttt{MR-GFT+CT-LL} against \texttt{MR-GFT+PPM}, \texttt{MR-GFT} and \texttt{HEVC} is presented in Fig.\;\ref{fig:RD_Curve}.
The proposed \texttt{MR-GFT+CT-SSDD} achieves promising bit rate reduction over a wide range of PSNR, and the proposed \texttt{MR-GFT+CT-LL} also outperforms other three schemes.
On average, we achieve a bit rate reduction of 19.34\% over \texttt{HEVC}, 10.29\% over \texttt{MG-GFT}, 5.02\% over \texttt{MR-GFT+PPM} and 3.46\% over \texttt{MR-GFT+CT-LL}.

Compared to \texttt{MR-GFT}, the gain of the proposed two schemes comes from the more efficient edge coding schemes.
Table \ref{tab:SSDD_edge_bits} gives the detailed results of bits of coding edges among the total bits with different $QP$ of \texttt{Teddy}.
When $QP=37$, we save on average 14.84\% bits and 5.17\% bits for coding edges with the proposed \texttt{CT-SSDD} and \texttt{CT-LL} compared to \texttt{AEC} adopted in \texttt{MR-GFT}.
Note that the bits of coding residuals (difference between total bits and bits of coding edges) of \texttt{MR-GFT+CT-LL} and \texttt{MR-GFT} are the same, while they are a bit larger than that of \texttt{MR-GFT+CT-SSDD}.
Using \texttt{MR-GFT+CT-SSDD}, the depth image is augmented with fewer irregular edges before coding, resulting in more efficient directional intra prediction.

The bit rate reduction of \texttt{CT-SSDD} over \texttt{CT-LL} becomes larger when the $QP$ increases.
This is also reflected in the RD curve in Fig.\;\ref{fig:RD_Curve}.
As previously discussed, the strength of the approximation is controlled by $\lambda$.
When $QP$ is small, $\lambda$ is also small which makes the edges less likely be changed.
In other words, in high bit rate situation, it is unnecessary to approximate the edges, while more edges can be changed in the low bit rate situation.  

\begin{figure}[htb]

\begin{minipage}[b]{1\linewidth}
  \centering
  \centerline{\includegraphics[width=10cm]{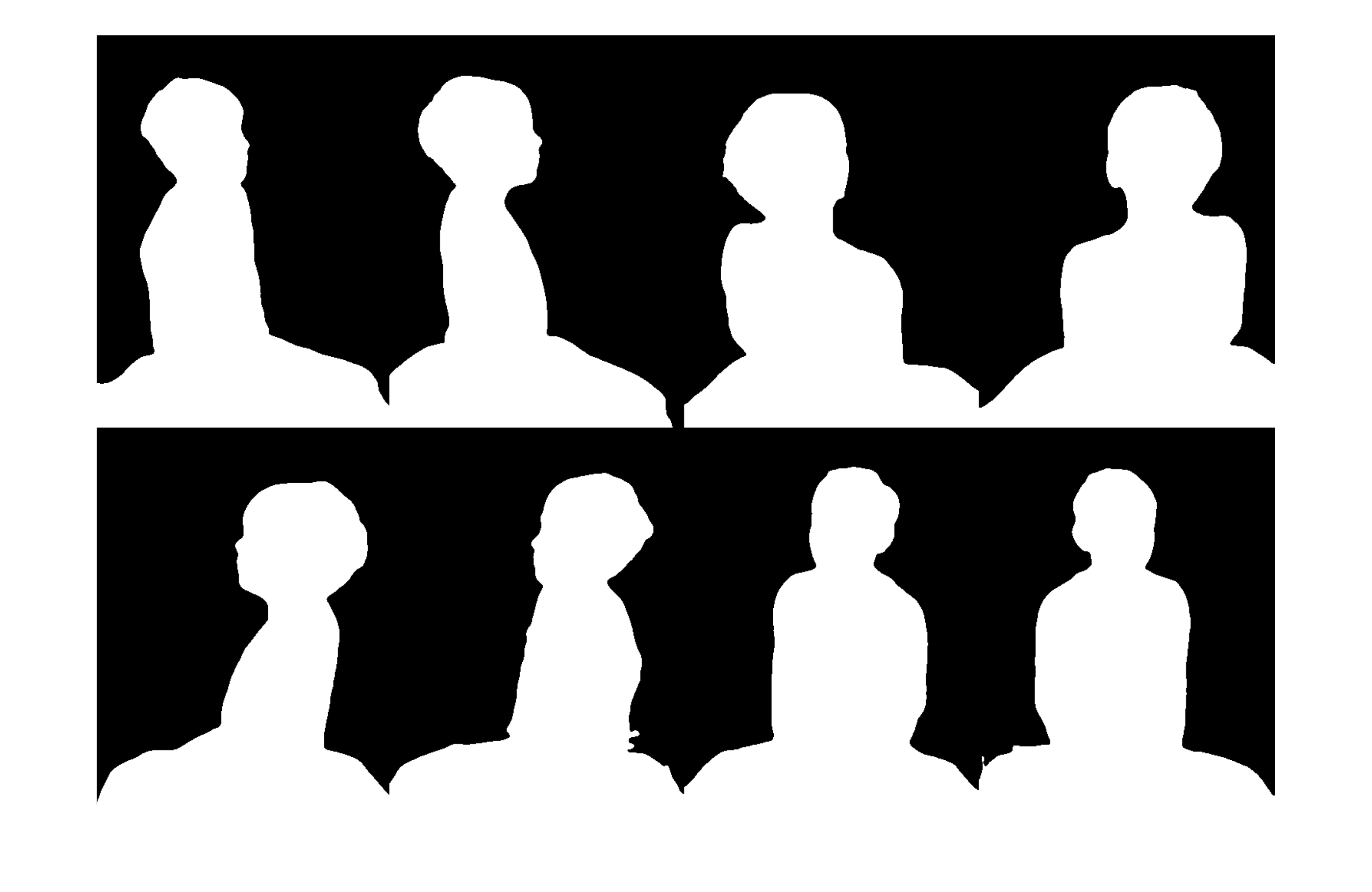}}
%  \vspace{1.5cm}
  \centerline{(a)}\medskip
\end{minipage}
\vfill
\begin{minipage}[b]{.48\linewidth}
  \centering
  \centerline{\includegraphics[width=4cm]{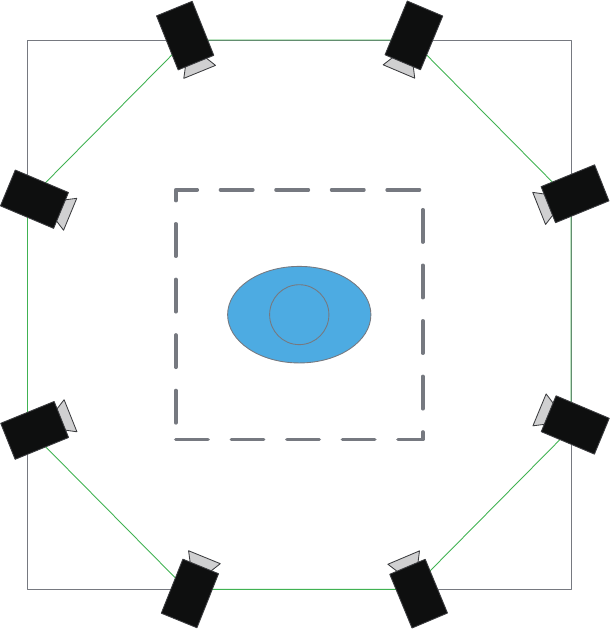}}
%  \vspace{1.5cm}
  \centerline{(b)}\medskip
\end{minipage}
\hfill
\begin{minipage}[b]{0.48\linewidth}
  \centering
  \centerline{\includegraphics[width=4cm]{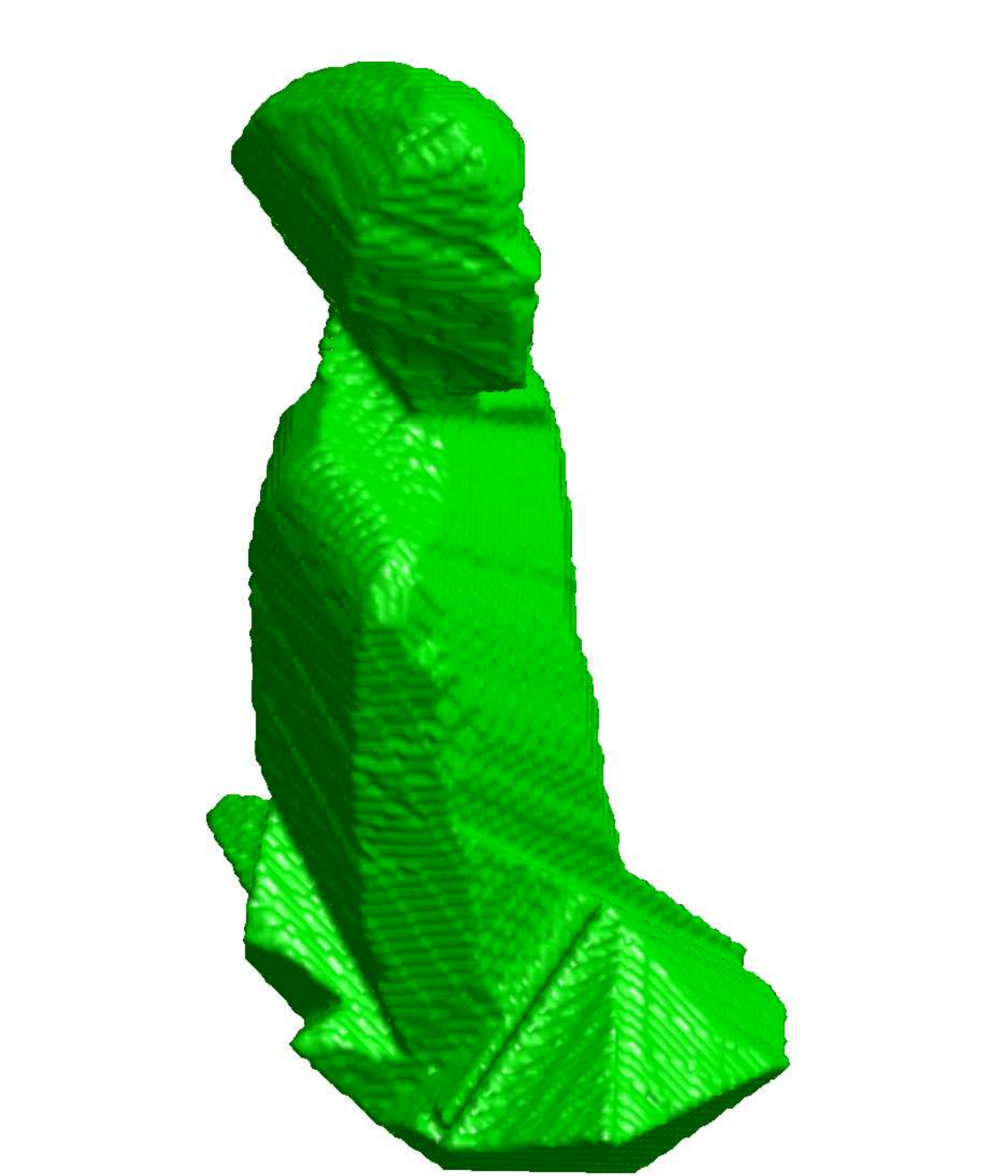}}
%  \vspace{1.5cm}
  \centerline{(c)}\medskip
\end{minipage}

\vspace{-0.3cm}
\caption{(a) Silhouettes of eight views for a single frame. (b) Top views of the capture rig layout showing 8 cameras and human subject within the cube. (c) One view of a model reconstructed from the silhouettes in (a).}
\label{fig:camera_setup}
\end{figure}

\subsection{Multiview Silhouettes Coding for 3D Reconstruction}
\label{ssec:lossy_MADD}

We test our proposed MADD based lossy contour coding \texttt{CT-MADD} on the multiview silhouette sequences from Microsoft Research.
The sequences are obtained using the equipment set up in \cite{loop2013real}.
The spatial resolution of the silhouettes is $384\times 512$.
Each sequence contains 10 frames, and eight views are taken for each frame.
Fig.\;\ref{fig:camera_setup}(a) shows an example of the silhouettes of eight views for a single frame.
The silhouettes were extracted from the images captured with an eight camera rig, as illustrated in Fig.\;\ref{fig:camera_setup}(b).
With the extracted silhouettes, the 3D model can be reconstructed as shown in Fig.\;\ref{fig:camera_setup}(c).
We coded four multiview silhouette sequences using \texttt{CT-MADD} and reconstructed the 3D models with the coded silhouettes.
The RD performance was evaluated, where distortion is the volume error between the 3D model reconstructed with the original silhouettes and that with the coded silhouettes.
 
The 3D model was reconstructed based on a volume intersection scheme \cite{martin1983volumetric}.
We projected each silhouette onto an initialized 3D volume and the final 3D model is the intersected volume of the eight projected silhouettes.

We compare the RD performance of \texttt{CT-MADD} against two lossy contour compression schemes: the operational rate-distortion optimal polygon-based shape coding (\texttt{ORD}) in \cite{katsaggelos1998mpeg} and the improved version based on an adaptive edge coding scheme (\texttt{EA-ORD}) in \cite{zhu2014adaptive}.
\texttt{ORD} and \texttt{EA-ORD} also adopt the MADD measure same as the proposed \texttt{CT-MADD}. For all three methods, $D_{\max}$ was set from 1 to 5 to test different distortion.

\begin{figure}[htb]

\begin{minipage}[b]{.33\linewidth}
  \centering
  \centerline{\includegraphics[width=2.8cm]{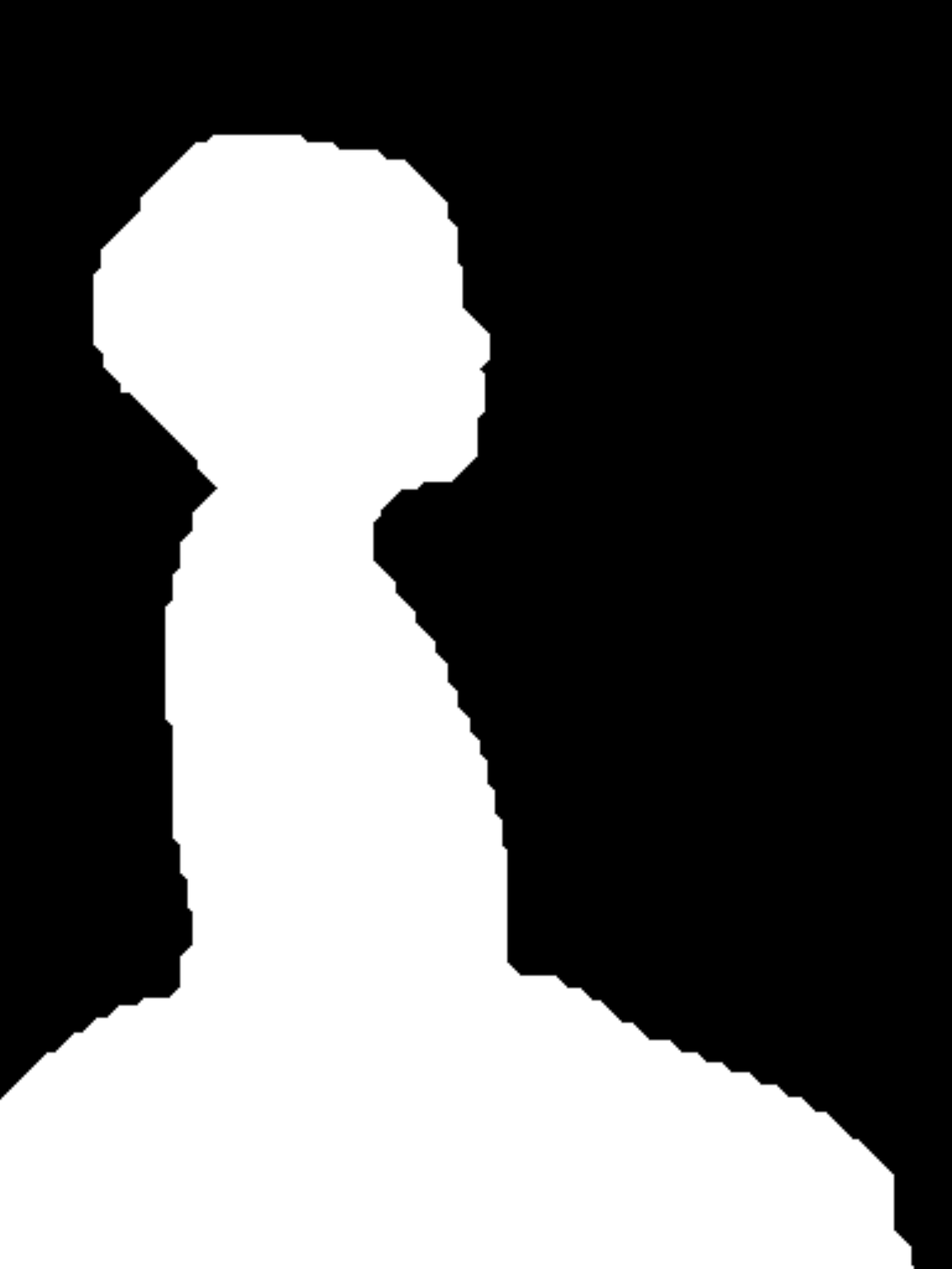}}
%  \vspace{1.5cm}
  \centerline{(a) 1429 bits}\medskip
\end{minipage}
\hfill
\begin{minipage}[b]{0.32\linewidth}
  \centering
  \centerline{\includegraphics[width=2.8cm]{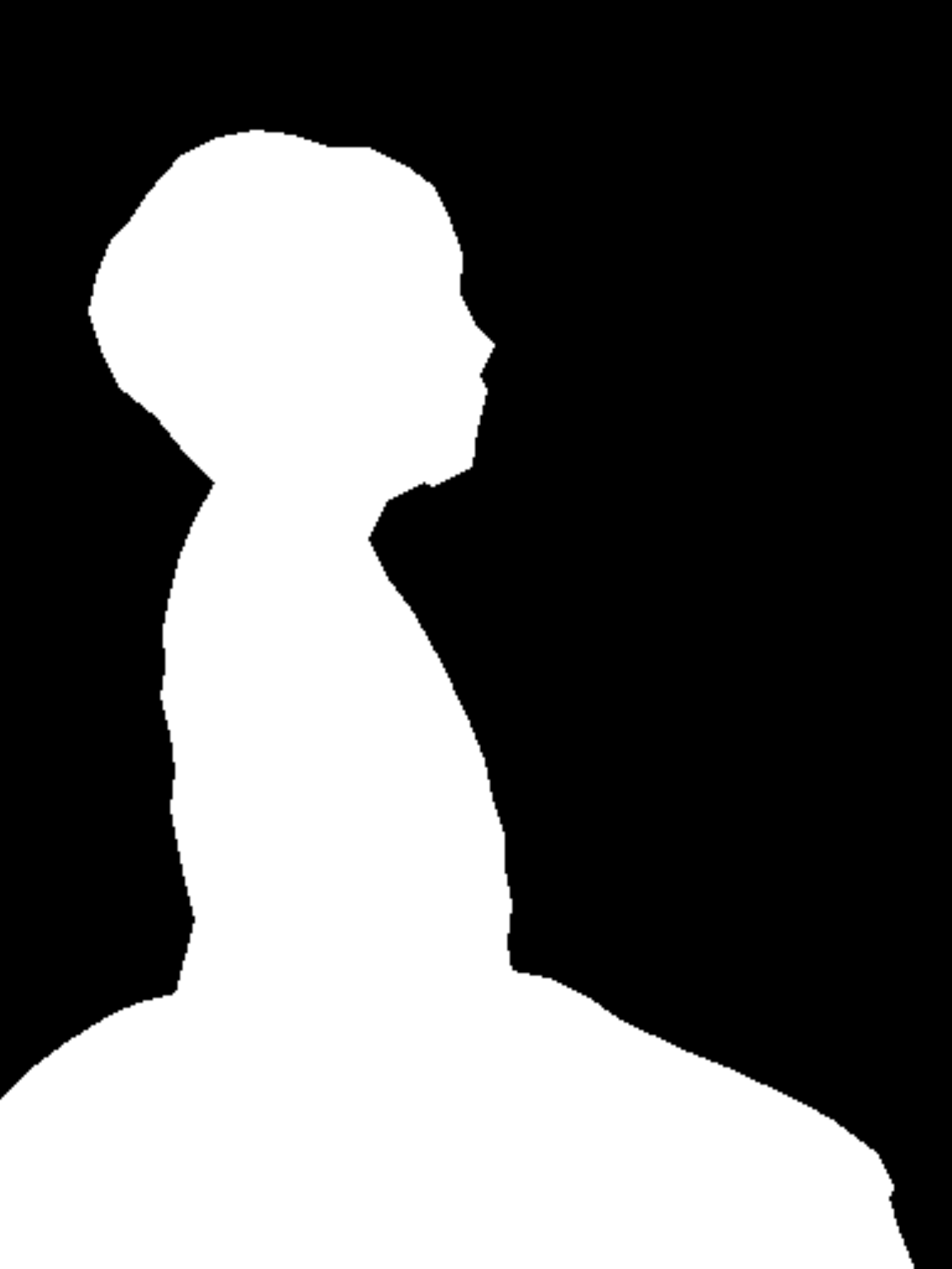}}
%  \vspace{1.5cm}
  \centerline{(b) 1179 bits}\medskip
\end{minipage}
\hfill
\begin{minipage}[b]{0.32\linewidth}
  \centering
  \centerline{\includegraphics[width=2.8cm]{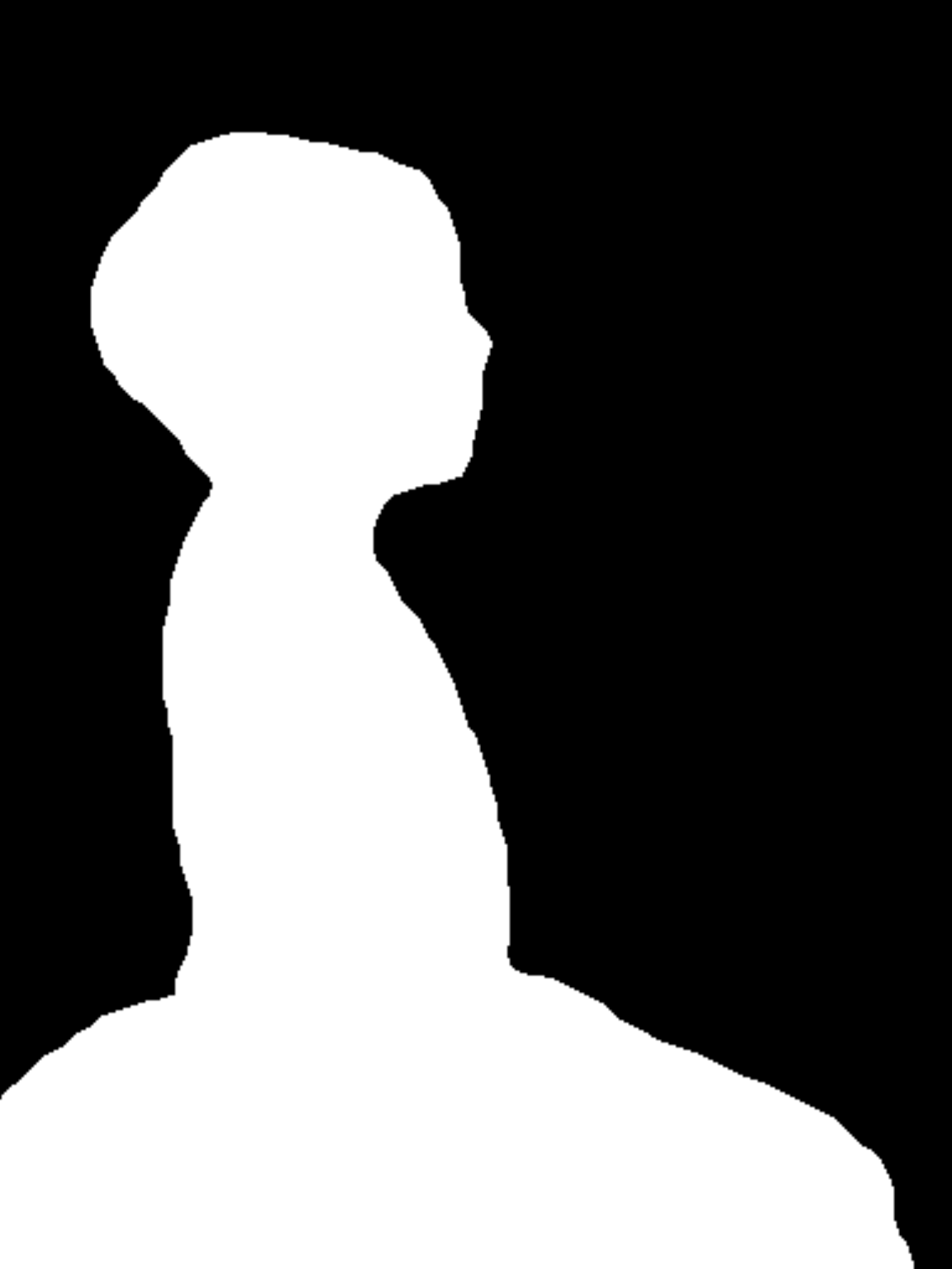}}
%  \vspace{1.5cm}
  \centerline{(c) 664 bits}\medskip
\end{minipage}

\begin{minipage}[b]{.32\linewidth}
  \centering
  \centerline{\includegraphics[width=2.8cm]{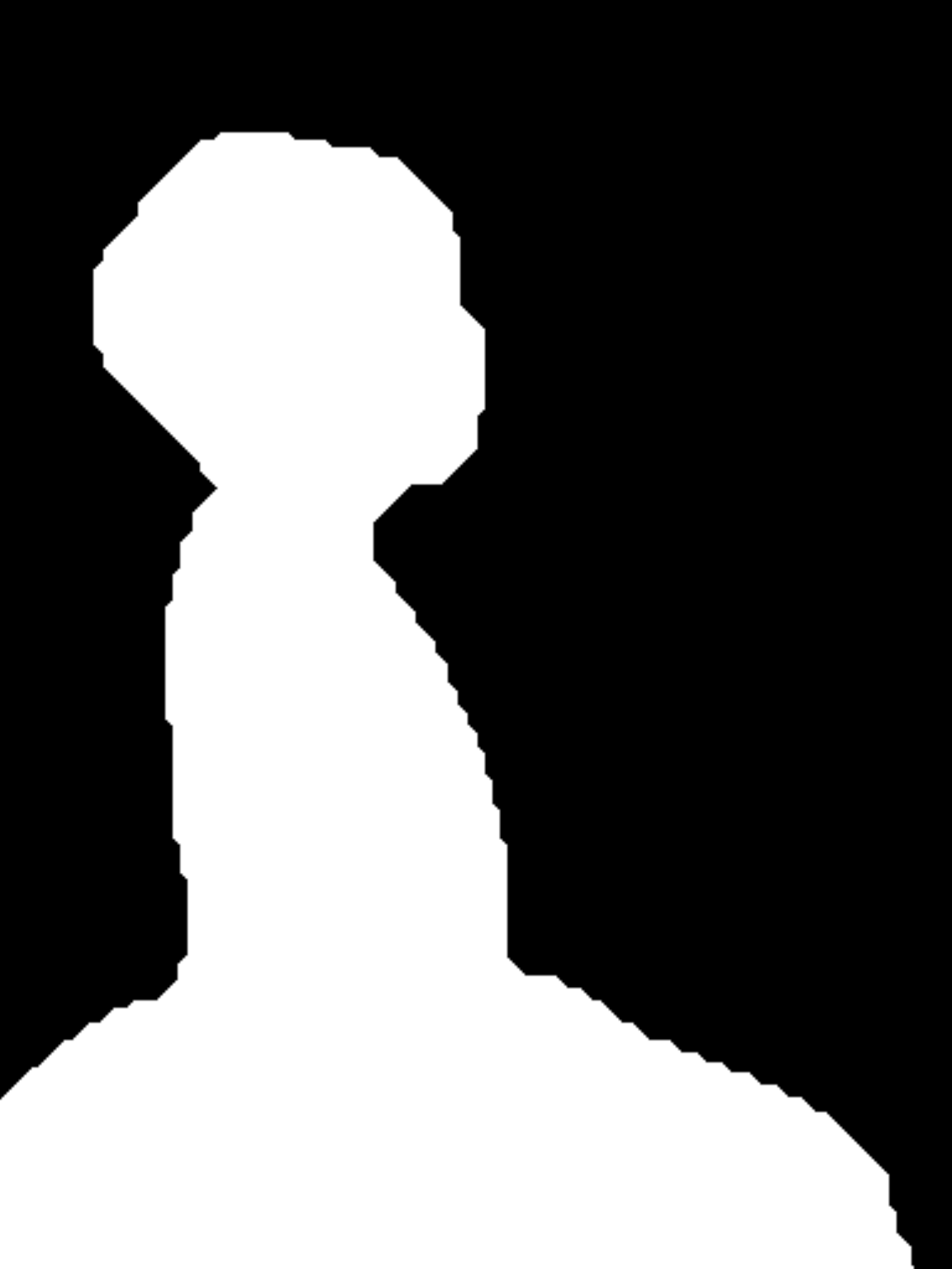}}
%  \vspace{1.5cm}
  \centerline{(d) 1376 bits}\medskip
\end{minipage}
\hfill
\begin{minipage}[b]{0.32\linewidth}
  \centering
  \centerline{\includegraphics[width=2.8cm]{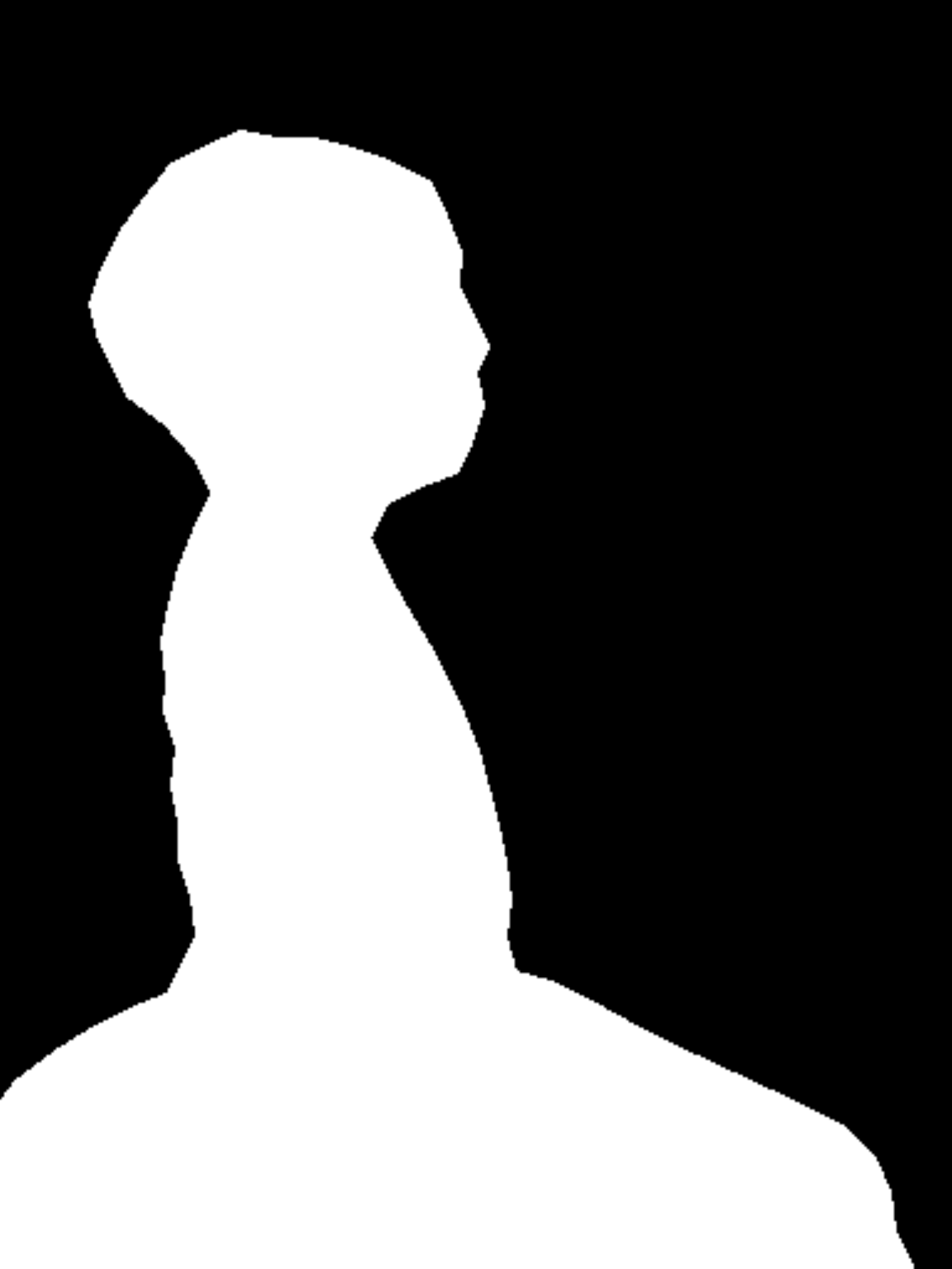}}
%  \vspace{1.5cm}
  \centerline{(e) 1128 bits}\medskip
\end{minipage}
\hfill
\begin{minipage}[b]{0.32\linewidth}
  \centering
  \centerline{\includegraphics[width=2.8cm]{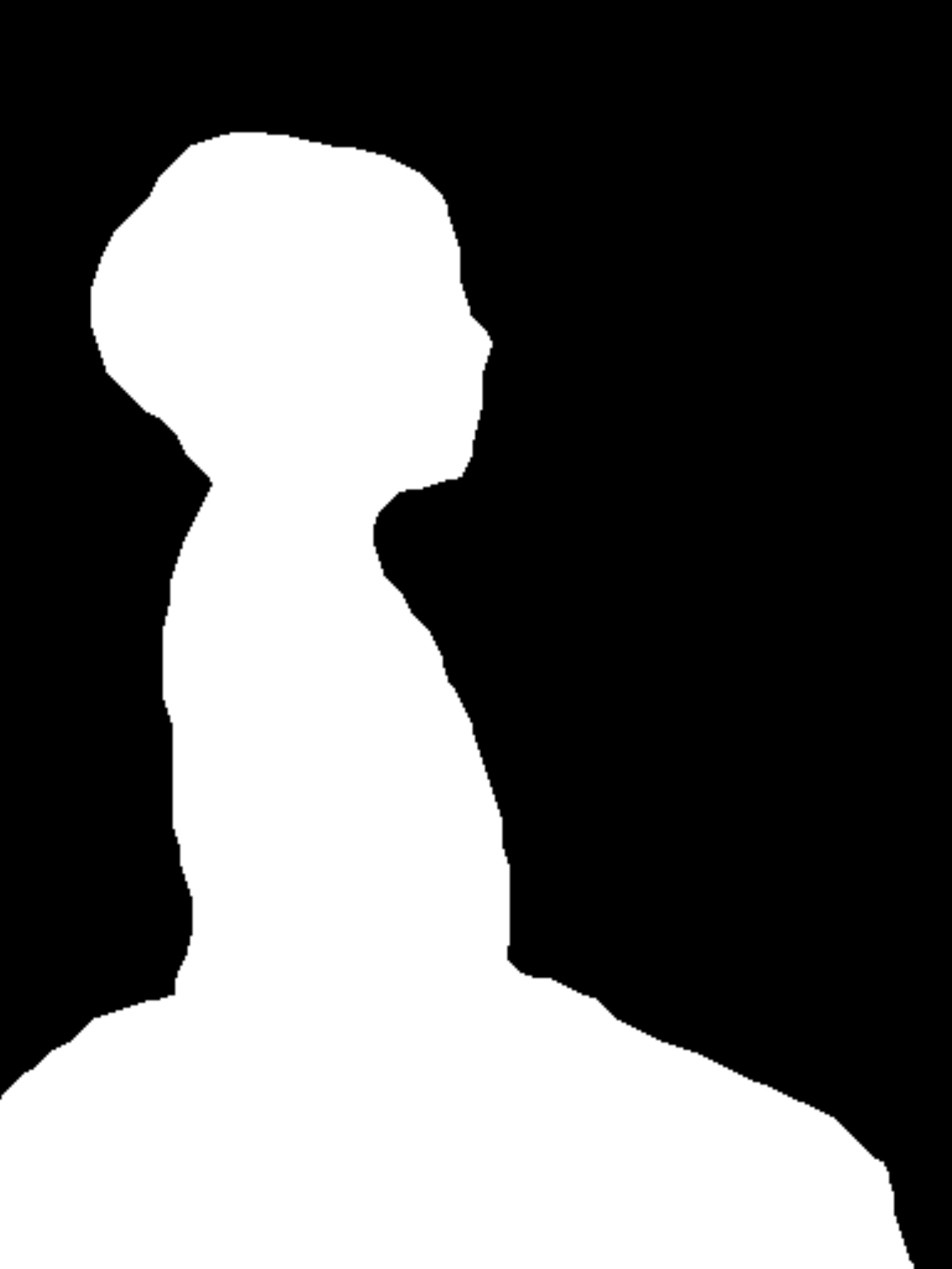}}
%  \vspace{1.5cm}
  \centerline{(f) 646 bits}\medskip
\end{minipage}

\vspace{-0.3cm}
\caption{Examples of silhouette approximation of different methods. The original silhouette is the second view in Fig.\;\ref{fig:camera_setup}(a). Left column: \texttt{ORD}. Middle column: \texttt{EA-ORD}. Right column: \texttt{CT-MADD}. (a)\;$\sim$\;(c) $D_{\max}=1$. (d)\;$\sim$\;(f) $D_{\max}=3$. }
\label{fig:silhouette_approximation}
\end{figure}

\subsubsection{Comparison of Contour Approximation}

Fig.\;\ref{fig:silhouette_approximation} shows one view of silhouette approximation using different methods with $D_{\max}=1$ and $D_{\max}=3$.
With the same $D_{\max}$, \texttt{CT-MADD} consumes the fewest bits to code the silhouette.
As for subjective quality, the silhouettes approximated by \texttt{CT-MADD} are most similar visually to the original silhouettes, while the results by \texttt{ORD} and \texttt{EA-ORD} contain lots of staircase shapes. 

In both \texttt{ORD} and \texttt{EA-ORD}, only some contour pixels are selected for coding.
The approximated silhouette was constructed by connecting these selected contour pixels in order, making the result unnatural with too many staircase lines.  

\begin{figure}[htb]

\begin{minipage}[b]{.48\linewidth}
  \centering
  \centerline{\includegraphics[width=4.5cm]{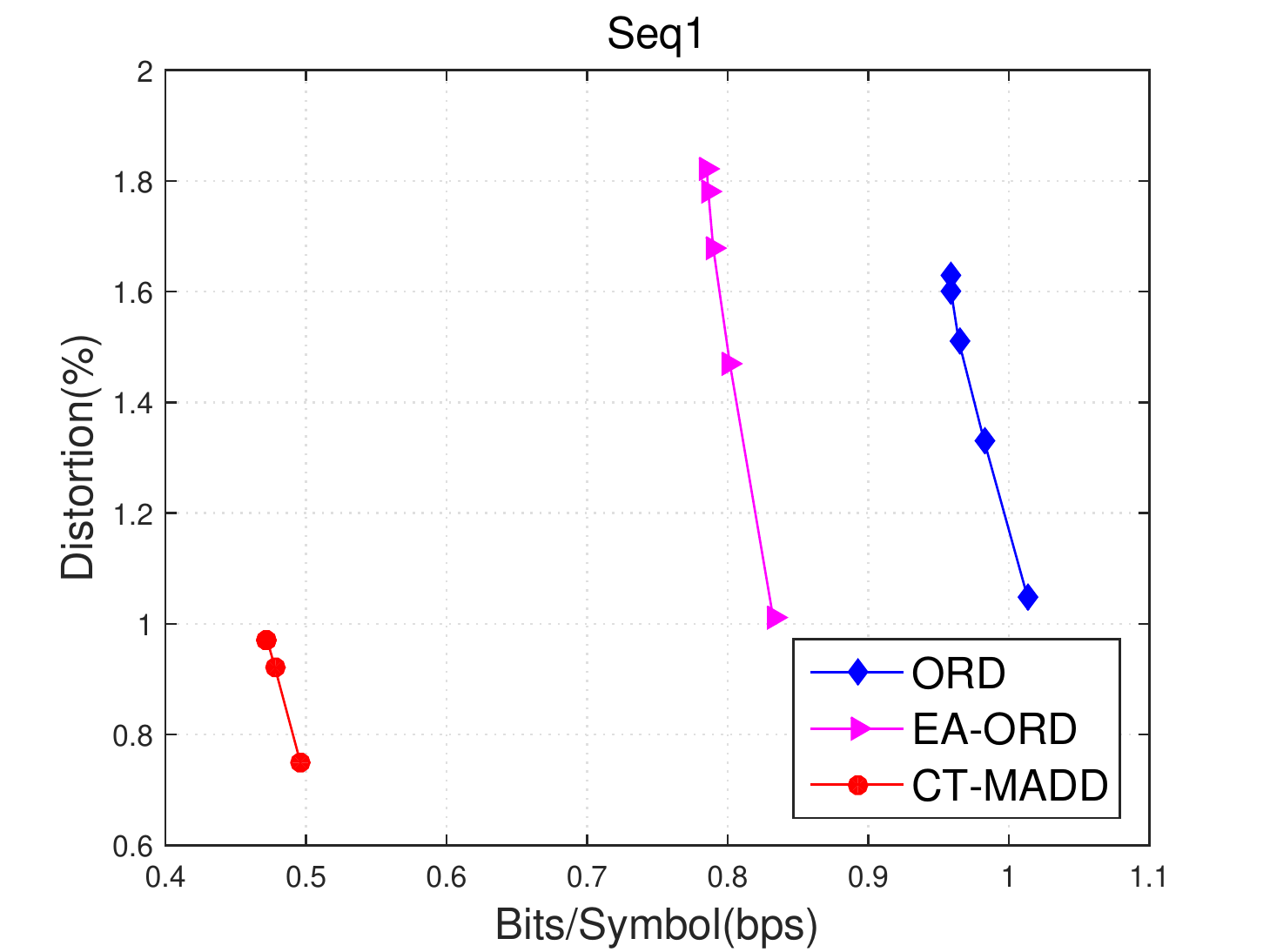}}
%  \vspace{1.5cm}
  \centerline{(a)}\medskip
\end{minipage}
\hfill
\begin{minipage}[b]{0.48\linewidth}
  \centering
  \centerline{\includegraphics[width=4.5cm]{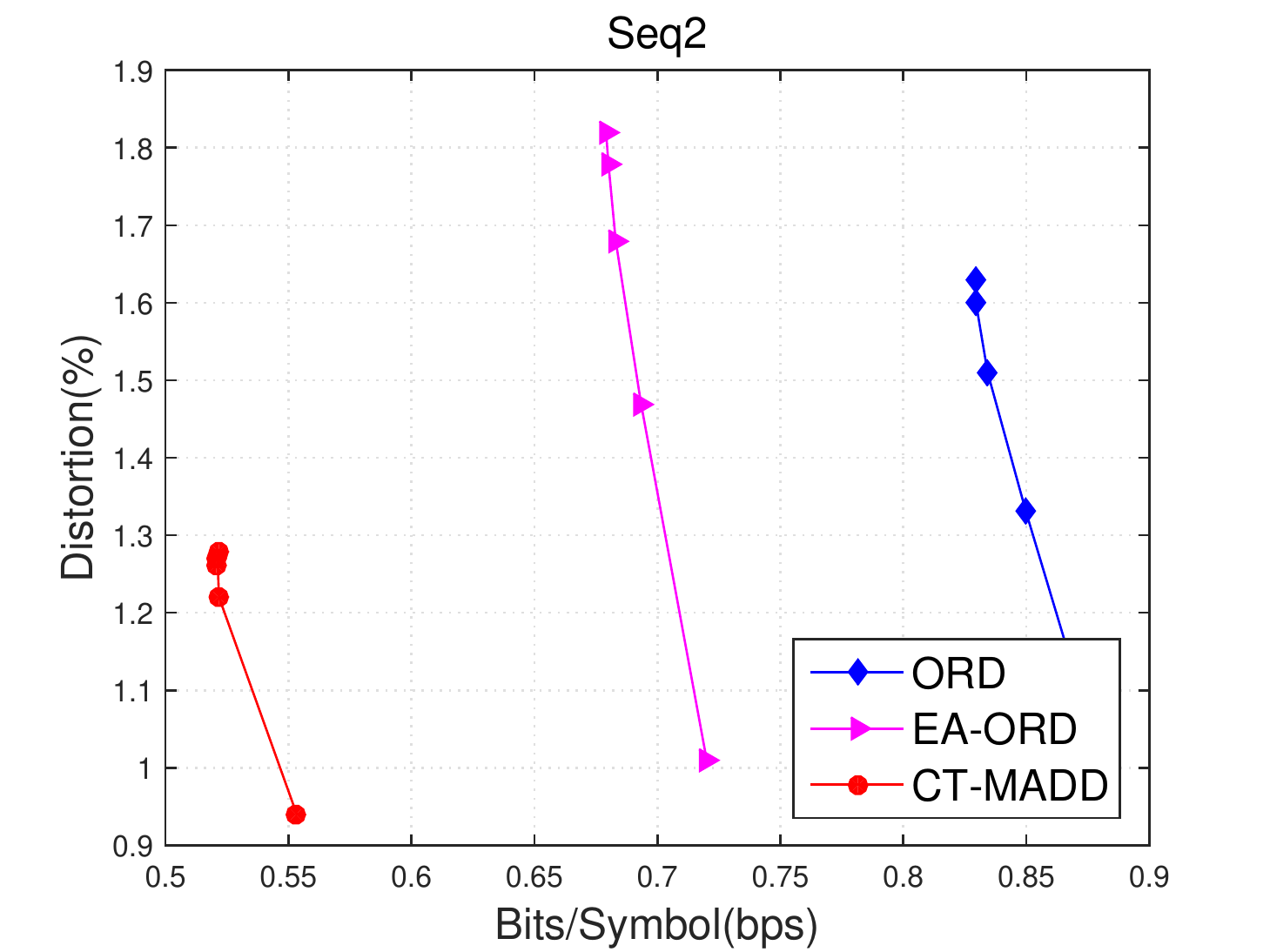}}
%  \vspace{1.5cm}
  \centerline{(b)}\medskip
\end{minipage}
\vfill
\begin{minipage}[b]{.48\linewidth}
  \centering
  \centerline{\includegraphics[width=4.5cm]{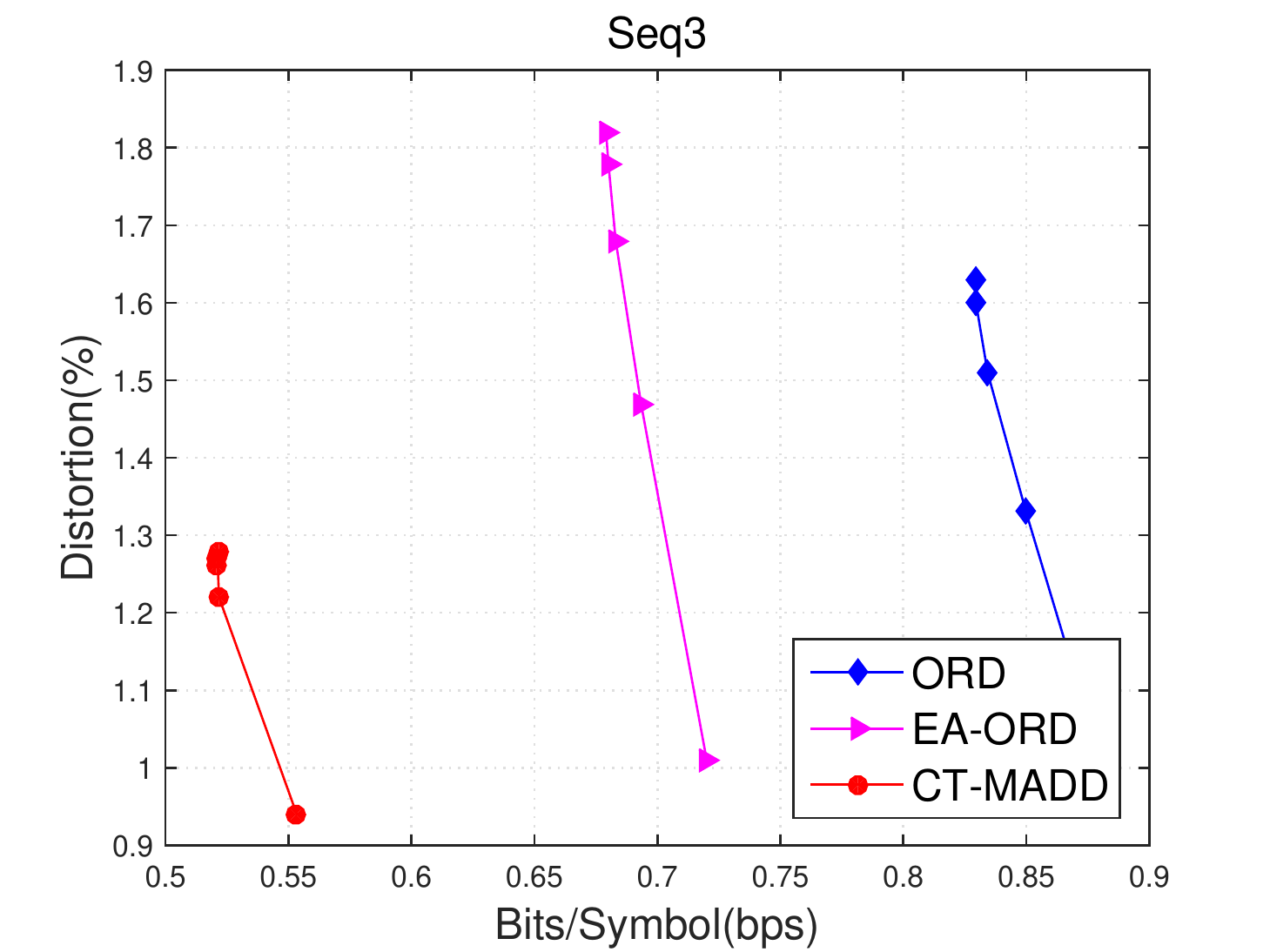}}
%  \vspace{1.5cm}
  \centerline{(c)}\medskip
\end{minipage}
\hfill
\begin{minipage}[b]{0.48\linewidth}
  \centering
  \centerline{\includegraphics[width=4.5cm]{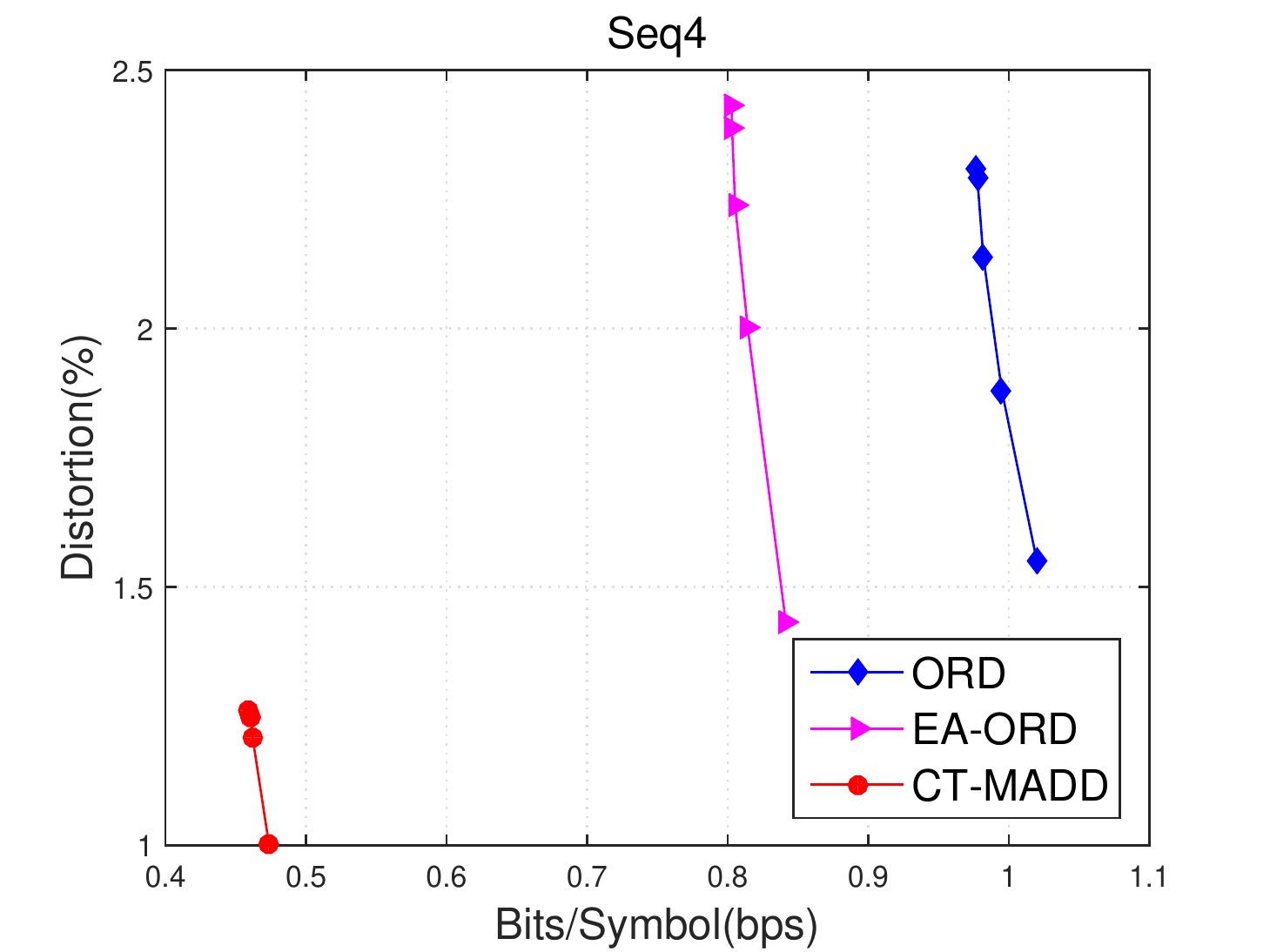}}
%  \vspace{1.5cm}
  \centerline{(d)}\medskip
\end{minipage}

\vspace{-0.3cm}
\caption{RD performance comparison among different compression schemes for multiview silhouettes coding.}
\label{fig:RD_Curve_Silhouettes}
\end{figure}

\subsubsection{Evaluation of RD Performance}

The RD performance is shown in Fig.\;\ref{fig:RD_Curve_Silhouettes}.
We observe that \texttt{CT-MADD} outperforms \texttt{ORD} and \texttt{EA-ORD} significantly both in bit rate and distortion.
Specifically, at the same $D_{\max}$, \texttt{CT-MADD} has an average bitrate reduction of 50\% and 38\% and an average of distortion reduction of 31\% and 43\% compared to \texttt{ORD} and \texttt{EA-ORD}.

In \texttt{ORD} and \texttt{EA-ORD}, the differences between two selected contour pixels (vertices) are coded using unary code.
Larger differences will consume more bits.
Although \texttt{CT-MADD} codes more number of contour pixels, the accurately estimated conditional probabilities enable our contour pixels to be coded much more efficiently than \texttt{ORD} and \texttt{EA-ORD}.

%% file: conclude.tex
We investigate the problem of lossless and lossy coding of image contours, focusing on the case when the sizes of training datasets $\mathcal{X}$ are limited. 
To avoid overfitting, we propose a \textit{maximum a posteriori} (MAP) formulation to compute an optimal variable-length context tree (VCT) $\mathcal{T}$---one that has both large likelihood $P(\mathcal{X}|\mathcal{T})$ that can explain observed data $\mathcal{X}$, and large prior $P(\mathcal{T})$ stating that contours are more likely straight than curvy.
For the lossy case, we develop dynamic programming (DP) algorithms that approximate a detected contour by trading off coding rate with two different notions of distortion.
The complexity of the DP algorithms can be reduced via compact table entry indexing using a total suffix tree (TST) derived from VCT $\mathcal{T}$. 
Experimental results show that our proposed lossless and lossy algorithms outperform state-of-the-art coding schemes consistently for both small and large training datasets.